\documentclass[11 pt]{article}
\usepackage[mathscr]{eucal}
\usepackage{amssymb,latexsym}
\usepackage{verbatim}

\usepackage{amsmath}
\usepackage{amsthm}
\usepackage{enumerate}
\usepackage{authblk}
\usepackage{color}
\usepackage{multicol}
\usepackage{tikz}
\usepackage{caption}
\usepackage{cite}
\usepackage[makeroom]{cancel}
\usepackage{dsfont}
\usepackage[nottoc,notlot,notlof]{tocbibind}

\newlength{\bibitemsep}
\newlength{\bibparskip}
\let\oldthebibliography\thebibliography
\renewcommand\thebibliography[1]{%
  \oldthebibliography{#1}%
  \setlength{\parskip}{\bibitemsep}%
  \setlength{\itemsep}{\bibparskip}%
}
\usepackage{hyperref}
\hypersetup{
    colorlinks=true,
    linkcolor=blue,
    filecolor=magenta,      
    urlcolor=cyan,
    citecolor=black,
}
\usepackage{url}

\newtheorem{thm}{Theorem}[section]
\newtheorem{lem}[thm]{Lemma}
\newtheorem{prop}[thm]{Proposition}
\newtheorem{rem}[thm]{Remark}

\newcommand\calM{{\mathcal{M}}}

\renewcommand\l{\lambda}

\newcommand\wt{\widetilde}

\newcommand{\alphaS}{\alpha_{\mbox{\tiny{S}}}}
\newcommand{\muB}{\mu_{\mbox{\tiny{B}}}}

\newcommand\Si{\Sigma}

\newcommand\s{\sigma}

\newcommand\e{\varepsilon}

\renewcommand\l{\lambda}
\newcommand\g{\gamma}

\renewcommand\a{\alpha}

\newcommand\beq{\begin{equation}}
\newcommand\eeq{\end{equation}}
\newcommand\ben{\begin{enumerate}}
\newcommand\een{\end{enumerate}}
\newcommand\bit{\begin{itemize}}
\newcommand\eit{\end{itemize}}

\newcommand{\La}{\Lambda}

\newcommand{\ov}{\overline}

\newcommand{\pd}{\partial}

\newcommand{\mc}{\mathcal}

\newcommand{\Z}{\mathbb{Z}}
\newcommand{\C}{\mathbb{C}}

\newcommand{\half}{\frac{1}{2}}

\newcommand{\bna}{\begin{eqnarray}}
\newcommand{\ena}{\end{eqnarray}}

\newcommand{\Cset}{\mathbb{C}}
\newcommand{\Nset}{\mathbb{N}}
\newcommand{\Rset}{\mathbb{R}}
\newcommand{\RR}{\mathbb{R}}

\newcommand{\Zset}{\mathbb{Z}}
 
\newcommand{\mEL}{m_{\mathrm{e}}}
\newcommand{\mPR}{m_{\mathrm{p}}} 

\def\undertilde#1{\mathord{\vtop{\ialign{##\crcr
   $\hfil\displaystyle{#1}\hfil$\crcr\noalign{\kern1.5pt\nointerlineskip}
   $\hfil\tilde{}\hfil$\crcr\noalign{\kern1.5pt}}}}}

\newcounter{mnotecount}

\title{On the discrete Dirac spectrum of general-relativistic hydrogenic ions with anomalous magnetic moment\footnote{\copyright The authors. Reproduction, in its entirety, is permitted for non-commercial purposes.}}

\author[1]{E. B. Kapengut}
\author[1]{M. K.-H. Kiessling}
\author[2]{E. Ling}
\author[1]{A. S. Tahvildar-Zadeh}

\affil[1]{Department of Mathematics, Rutgers University (New Brunswick)}
\affil[2]{Copenhagen Centre for Geometry and Topology (GeoTop),
\linebreak
Department of Mathematical Sciences, University of Copenhagen, Denmark}

\begin{document}

\maketitle

\begin{abstract}\noindent
The Reissner--Weyl--Nordstr\"om (RWN) spacetime of a point nucleus features a naked singularity for the empirically known nuclear charges $Ze$ and masses $M = A(Z,N)\mPR$, where $\mPR$ is the proton mass and $A(Z,N)\approx Z+N$ the atomic mass number, with $Z$ the number of protons and $N$ the number of neutrons in the nucleus.
The Dirac Hamiltonian for a test electron with mass $\mEL$, charge $-e$, and anomalous magnetic moment $\mu_a (\approx - \frac{1}{4\pi}\frac{e^3}{\mEL c^2})$ in the electrostatic RWN spacetime of such a `naked point nucleus' is known to be essentially self-adjoint, with a spectrum that consists of the union of the essential spectrum $(-\infty,-\mEL c^2]\cup[\mEL c^2, \infty)$ and a discrete spectrum of infinitely many eigenvalues in the gap $(-\mEL c^2,\mEL c^2)$, having $\mEL c^2$ as accumulation point.
In this paper the discrete spectrum is characterized in detail for the first time, {for all $Z\leq 45$ and $A$ that cover all known isotopes.
The eigenvalues are mapped one-to-one to those of the traditional Dirac hydrogen spectrum.
Numerical evaluations that go beyond $Z=45$ into the realm of not-yet-produced hydrogenic ions are presented, too.
A list of challenging open problems concludes this publication.}
\end{abstract}



\section{Introduction}
Since hydrogenic ions, with hydrogen here counted as the neutral special case, are the simplest atomic systems that are empirically stable in isolation, their study has played a prominent role in the development of quantum theory, and we are confident that this role is not over yet.
In particular, while the spectacular quantitative agreement of the theoretically computed energy spectrum with empirically available data is often mentioned in support of the well-known claim that QED is one of the most precise physical theories ever discovered \cite{QED}, viewed from a mathematically rigorous and physically principled standpoint this claim is just hype. 
This harsh assessment is forced upon one if one recalls how the theoretical computation of the hydrogenic spectrum has been done: by symbolically using first-order perturbation theory to compute corrections to the non-relativistic Bohr spectrum of the relative Pauli Hamiltonian of a hydrogenic ion with Coulomb interactions between the nucleus and the electron, where this relative Pauli Hamiltonian is the Friedrichs extension of the pertinent mathematical expression $-\frac{\hbar^2}{2\mu}\Delta - Z\frac{e^2}{r}$ defined on its minimal domain; here, $Z\in\Nset$ is the usual atomic number of the hydrogenic ion, $e$ the elementary charge, $\hbar$ is Planck's constant divided by $2\pi$, and $\mu=\mEL M/(\mEL +M)$ is the reduced mass of the hydrogenic ion (with $\mEL$ the electron's and $M$ the nucleus' mass).
The first-order corrections include the computation of the quantum-mechanical expected values, as per Born's rule, of relativistic mass and retardation effects, and the spin-orbit coupling, nuclear contributions due to their size and magnetic moment, and quantum field-theoretical effects such as Lamb shift and the contributions to the anomalous magnetic moment, the latter through evaluation of Feynman diagrams of small order in $\alphaS$ and $\ln \alphaS$, where
\begin{equation}\label{eq:alphaS}
\alpha_{\mbox{\tiny{S}}} \,=\, \frac{e^2}{\hbar c} \approx \frac{1}{137.036}
\end{equation} 
is Sommerfeld's fine structure constant, with $c$ the speed of light in vacuum.
These perturbative computations of the hydrogenic spectrum do not constitute a rigorous approximate evaluation of a non-perturbatively defined version of QED with an external charge (representing the nucleus), which would be required to vindicate the above cited claim of QED's supreme status.
One might think that the problem is caused by the fact that a proper treatment of a nucleus is not possible within QED, but the situation is not any better if one wants to extract the hydrogenic energy spectrum from the standard model of particle physics.
Moreover, positronium, though unstable, is the closest analog of a hydrogenic ion that should be accurately describable entirely within QED, but even the positronium spectrum has been computed so far only in the same perturbative manner as the hydrogenic ion spectrum \cite{Czarnecki}, mentioned above.
Clearly, much work still needs to be done to reach a conceptually satisfactory special-relativistic state of affairs.

he reader will have noticed that corrections due to gravitational effects between nucleus and electron have not been taken into account in these hydrogenic spectral computations. 
The rationale for this omission is a ratio of less than $3\times 10^{-39}$ of the non-relativistic gravitational Newton coupling constant compared to the electrical Coulomb coupling constant, which indicates that gravitational corrections are not only quantitatively unimportant but immeasurably tiny. 
However, conceptually speaking, it would be more satisfying if the quantitative insignificance of internal gravitational effects on the hydrogenic spectra would be demonstrated by a general-relativistic treatment (here we paraphrased Wereide \cite{Wereide}; cf. \cite{KieTaZaTopJMP}).

In this paper we report our progress made on the non-perturbative characterization and computation of the general-relativistic quantum-{\it mechanical} energy spectrum of hydrogenic ions. 
To the best of our knowledge, a rigorous formulation as a relativistic two-body problem is an outstanding open problem.
We therefore will employ the standard Born--Oppenheimer approximation that treats the electron as a test particle in the electromagnetic and gravitational fields of a given, fixed nucleus.
This approximation is expected to get more accurate with increasing $Z$.
The structure of the nucleus will be ignored too and replaced by a point particle, as also done in most rigorous works on the relativistic problem.
This approximation is expected to get worse with increasing $Z$, though.
Although these approximations will no doubt have a much much larger quantitative effect on the computation of hydrogenic spectra than the step from an exactly solvable special- to a mathematically challenging general-relativistic treatment, this should be of no concern because our work is meant to be of a conceptual nature and not aimed at some ultra-fine quantitative comparison of theoretical and empirical spectral values.
 
We now prepare the ground for the statement of our new contributions by briefly recalling the state of affairs, beginning with the special-relativistic limit where one ``switches gravity off'' by formally setting Newton's constant of universal gravitation $G\to 0$.
In the absence of gravity the default Dirac operator of hydrogenic ions is
\begin{equation}\label{eq:HdiracSR}
    H = -i \hbar c \vec{\boldsymbol{\alpha}}\cdot \nabla + \mEL c^2 \boldsymbol{\beta} - Ze^2\frac1r \boldsymbol{1},
\end{equation}
acting on four-component complex bi-spinor wave functions $\Psi$ on the minimal domain $\big(C^\infty_c (\RR^3\backslash\{0\})\big)^4\subset \big(L^2(\RR^3)\big)^4$.
As shown first by Weidmann \cite{Weidmann71}, when treating $Z$ as a positive real parameter, this special-relativistic Dirac operator for hydrogenic ions with purely electrical Coulomb interactions is essentially self-adjoint on $\big(C^\infty_c(\Rset^3\backslash\{0\})\big)^4$ for $Z \leq \sqrt{3}/2\alphaS (\approx 118.677)$; cf. \cite{Thaller}.
This covers all empirically known nuclei, not only those that occur in nature ($Z\leq 92$) but also all those produced artificially in nuclear reactors or heavy ion collisions ($Z\leq 118$).
Its discrete energy spectrum coincides with the well-known Sommerfeld fine structure formula
\begin{equation}\label{SommerfeldFS}
     E_{n,k}^{\rm Dirac} = \frac{\mEL c^2}{\sqrt{ 1 + \frac{Z^2 \alphaS^2}{\left(n-|k|+\sqrt{k^2-Z^2\alphaS^2}\right)^2}}},
\end{equation}
with $n\in \Nset$ and $k \in\{ -n,\dots,-1,1,\dots,n-1\}$, as shown first by Darwin \cite{DarwinDIRACspec} and Gordon \cite{GordonDIRACspec}, and reproduced many times since with various techniques (cf. \cite{ThallerBOOK}, \cite{GMRonQED}).

Sommerfeld's fine structure formula (\ref{SommerfeldFS}) is well-defined also for $Z\in\{119,...,137\}$ when (\ref{eq:HdiracSR}) is not essentially self-adjoint on its minimal domain, but does have a \emph{distinguished} self-adjoint extension for all positive $Z\leq \frac{1}{\alphaS}$ ($\approx 137.036$), defined by allowing $Z\in\Cset$ and demanding analyticity in $Z$ in the neighborhood of $Z=\sqrt{3}/2\alphaS$, as shown by Narnhofer \cite{Narnhofer}.
By analyticity it follows that (\ref{SommerfeldFS}) gives the discrete spectrum of this distinguished self-adjoint extension. 
It has been pointed out in \cite{GMRonQED} that this analytic extension is also the only one for which all bound states do have finite quantum expected values separately for kinetic and for potential energy; cf. Schmincke's thesis \cite{Schmincke}.
 
For integers $Z>137$ the Dirac operator (\ref{eq:HdiracSR}), still well-defined on its minimal domain, has infinitely many self-adjoint extensions and none is known as distinguished.  

As noted first by Behncke \cite{Behncke}, and explored subsequently by Arnold--Kalf--Schneider \cite{AKSch} and by Gesztezy--Simon--Thaller \cite{GesSimTha} (see also \cite{ThallerBOOK}), the lack of essential self-adjointness of (\ref{eq:HdiracSR}) for when $Z> 118$ (or $Z> \sqrt{3}/2\alphaS$ if $Z>0$ is treated as real) can be resolved by taking the anomalous magnetic moment $\mu_a$ of the electron into account.
This means to replace (\ref{eq:HdiracSR}) by 
\begin{equation}\label{eq:HdiracSRamm}
H = -i \hbar c \vec{\boldsymbol{\alpha}}\cdot \nabla + \mEL c^2 \boldsymbol{\beta} -Ze^2\frac1r \boldsymbol{1} + i \mu_a Ze \boldsymbol{\beta} \vec{\boldsymbol{\alpha}} \cdot\nabla\frac1r,
\end{equation}
with $\mu_a <0$ \cite{FanETal}. 
The operator (\ref{eq:HdiracSRamm}) on its minimal domain is essentially self-adjoint for all $Z>0$. 

Since the anomalous magnetic moment of the electron is conventionally explained as a quantum-electrodynamical effect, it could seem that a non-rigorous perturbative quantum field-theoretical calculation \cite{CetAL} is here invoked to rescue quantum mechanics from one of its dilemmas, reminiscent of the patchwork approach to the computation of the hydrogen spectrum that we have criticized above. 
But now consider this: In magnitude, the leading order contribution to the anomalous magnetic moment of the electron, according to expansion of QED in powers of Sommerfeld's $\alphaS$ and its logarithm, is the product of the Bohr magneton $\muB= \frac{1}{4\pi}\frac{he}{\mEL c}$ and the factor $\frac{1}{2\pi}\alphaS$, with $\alphaS$ given in (\ref{eq:alphaS}).
Thus the anomalous magnetic moment of the electron in quite accurate approximation reads $\mu_a \approx -\mu_{\rm cl}$, with
 \begin{equation}\label{eq:muCL}
\mu_{\rm cl} : = \frac{1}{4\pi}\frac{e^3}{\mEL c^2}  ,   
 \end{equation}
which is independent of $\hbar$.
As such (\ref{eq:muCL}) has an explanation as a classical physics quantity, named the \emph{classical magnetic moment of the electron} in \cite{KieTaZaTopJMP}.
Therefore the Dirac operator (\ref{eq:HdiracSRamm}) can be viewed as obtained by quantizing a classical special-relativistic theory of an electron with charge $-e$ and anomalous magnetic moment of magnitude $\mu_{\rm cl}$, as expected for a properly quantum-mechanical theory of hydrogenic ions.

The next task is to characterize the spectrum of (\ref{eq:HdiracSRamm}).
An important rigorous result has been contributed by Kalf and Schmidt \cite{KalfSchmidt}, and completed by Schmidt \cite{Schmidt}. 
These authors noted that when $Z\leq 118$ (or rather $Z\leq \frac{\sqrt{3}}{2}\frac{1}{\alphaS}$), then (\ref{eq:HdiracSRamm}) converges in strong resolvent sense to (\ref{eq:HdiracSR}) in the limit $\mu_a\to 0$.
This alone does not imply convergence of the point spectra.
They then went on to prove that when $Z\leq 137$ (or rather $Z < \frac{1}{\alphaS}$), then the eigenvalues of the unique self-adjoint operator with anomalous magnetic moment converge to those of the distinguished self-adjoint extension of the one without.
Thus, if $|\mu_a|$ is small enough, then the spectral results will differ only ever so slightly.

But how small is ``small enough''?
In particular, is the physical value of $|\mu_a|$ given in (\ref{eq:muCL}) small enough?
In the physics literature one traditionally finds mostly non-rigorous perturbative treatments; cf. \cite{CetAL}.
A non-perturbative numerical evaluation of a handful of low-lying eigenvalues has been carried out by the Thallers, see \cite{ThallerBOOK} (section 7, and the dedication) and \cite{ThallerREVIEW}, though unfortunately without revealing the details of how the numerical evaluation has been accomplished.
Numerically they compared the electron's lowest eigenvalue of (\ref{eq:HdiracSR}) with that of (\ref{eq:HdiracSRamm}). 
Their eigenvalue plot (see Fig.~7.1 in \cite{ThallerBOOK}) suggests that these lowest eigenvalues must be barely distinguishable for all naturally occurring $Z$, while significant discrepancies become visible when $Z>118$; the lowest eigenvalue of (\ref{eq:HdiracSR}) terminates when $Z\nearrow 137.036$, while its counterpart for (\ref{eq:HdiracSRamm}) continues to exist until it meets the previously second-lowest eigenvalue at about $Z\approx 145$ with value $E\approx 0$. 
There a cross-over happens, and both eigenvalues continue to decrease until they meet the negative continuum roughly when $Z\approx 147$, resp. $Z\approx 160$.
We have not seen any confirmation of these numerical results by other authors.
This is one of the open issues that we address in our paper, as the special-relativistic spin-off of our main inquiry.
 
We thus come to the general-relativistic problem. 
As in the pioneering investigations by Wereide \cite{Wereide}, Vallarta \cite{Vallarta}, Cohen and Powers \cite{CohenPowers}, and Belgiorno et al. \cite{BMB}, as spacetime of the fixed point nucleus here we also take the Reissner--Weyl--Nordstr\"om (RWN) solution of Einstein's field equations, discovered independently in \cite{WeylRWN}, \cite{ReissnerRWN}, \cite{NordstromRWN}. 
The RWN spacetime is a static, spherically symmetric Lorentzian manifold with two real parameters: the ADM mass $M>0$, and the charge $Q$; for hydrogenic ions, one sets $Q = Ze > 0$, and $M$ is identified with the empirical mass of the nucleus.
Using the $(+, -,-,-)$ signature, the line element corresponding to the Lorentzian metric is
\begin{equation}\label{eq:StaticSO3metric}
ds^2 \,=\, f(r)^2  c^2dt^2 - \frac{1}{f(r)^2}dr^2 - r^2 d\Omega^2,
\end{equation}
where $d\Omega^2$ is the usual round metric on $\mathbb{S}^2$ and 
\begin{equation}\label{eq:RWNfSQR}
f(r)^2 \,=\, {1 - \frac{G}{c^2}\,\frac{2M}{r} + \frac{G}{c^4}\,\frac{Z^2e^2}{r^2}}\,. 
\end{equation}
All known nuclei are in the naked singularity sector, $GM^2 < Z^2e^2$; in fact, we have $GM^2 \ll Z^2e^2$ by more than thirty orders of magnitude.
In this case, $f(r)^2$ is strictly positive for all $r > 0$, and we stipulate $f(r)$ to be its positive square root. 
There is a curvature singularity at $r = 0$; consequently, the maximal-analytically extended manifold is  ${\calM}=\RR\times (0,\infty) \times \mathbb{S}^2$. 

The general-relativistic counterpart to Dirac's operator (\ref{eq:HdiracSR}) is the Dirac operator extracted from Dirac's equation for a point electron in the RWN spacetime of a point nucleus,
 \begin{equation}\label{eq:DiracNOammGR}
\tilde{\g}^\mu(i\hbar c\nabla_\mu -e A_\mu)\Psi - \mEL c^2 \Psi \,=\,0.
\end{equation}
Here, $\Psi \colon {\calM} \to \C^4$ is a bi-spinor field on the spacetime manifold,  $\nabla_\mu$ is the spin connection acting on bi-spinors, and $\tilde{\g}^\mu$ are the Dirac matrices that satisfy the anti-commutation relations with respect to the Lorentzian metric (\ref{eq:StaticSO3metric}), (\ref{eq:RWNfSQR}). 
The Dirac matrices generate the complexified Clifford algebra $\text{Cl}_{1,3}(\RR)_\mathbb{C}$ over each tangent space $T_p{\calM}$ on the manifold. 
Finally, $A = -\frac{Q}{r}cdt$ is the one-form of the electromagnetic potential of the RWN spacetime, which is purely electrostatic.
In the $(x^0 =ct, r, \theta, \phi)$ coordinates, the only nonzero component of $A$ is $A_0 = -\frac{Q}{r}$. 
The quantum-mechanical Dirac operator $H$ is now extracted from (\ref{eq:DiracNOammGR}) by rewriting this Dirac equation into the abstract Schr\"odinger format $i\hbar \partial_t \Psi = H \Psi$.

It came as a big surprise when in 1982 Cohen and Powers \cite{CohenPowers} proved that this general-relativistic Dirac operator for hydrogenic ions is not essentially self-adjoint on its minimal domain but has a continuously parameterized family of self-adjoint extensions.
They thus proved that, unlike Newton's gravity in the corresponding non-relativistic Schr\"odinger operator, Einstein's gravity is not a weak perturbation of Coulomb electricity in the Dirac problem for hydrogenic ions for all $Z<118$.

About 20 years later, Belgiorno, Martinelli, and Baldicchi \cite{BMB} showed that the replacement of (\ref{eq:DiracNOammGR}) with the Dirac equation that includes a term for the electron’s anomalous magnetic moment coupled to the gradient of the Coulomb field (we present this Dirac equation in the next section) restores essential self-adjointness --- though only if the magnitude of the anomalous magnetic moment is larger than the critical value $\mu_{\rm crit}^{}$ given by
\begin{equation}\label{eq:BMBmu}
 \mu_{\rm crit}^{} :=  \frac32\frac{\hbar\sqrt{G}}{c}.
\end{equation} 
Put differently, the essential self-adjointness of the special-relativistic Dirac operator for hydrogenic ions with any $Z>0$ when the electron has both charge and anomalous magnetic moment is not mathematically structurally stable against the ``switching on'' of Einstein's gravity, unless the anomalous magnetic moment in magnitude is larger than the critical value (\ref{eq:BMBmu}).

Comparison of (\ref{eq:muCL}) with (\ref{eq:BMBmu}) reveals that the empirical value of the anomalous magnetic moment is about $10^{18}$ times larger than the theoretical critical value. 
Therefore, essential self-adjointness of this general-relativistic Dirac operator for a physically relevant model of hydrogenic ions with empirically accurate anomalous magnetic moment of the electron has been established. 
This holds for all $Z>0$; though, for $Z>118$ we leave the empirically confirmed realm of nuclei, as noted earlier.
For this essentially self-adjoint general-relativistic Dirac operator, the authors in \cite{BMB} showed that the essential spectrum consists of the closure of the usual positive and negative absolutely continuous spectra outside the interval $[-\mEL c^2, \mEL c^2]$, and that this interval contains an infinite discrete spectrum.

\subsection{Informal statement of our main result}

The main purpose of the present paper is to supply a complete characterization of the discrete spectrum of the Dirac Hamiltonian for a point electron with larger-than-critical anomalous magnetic moment in the RWN spacetime of a point nucleus.
Our main result, Theorem \ref{thm: main}, states that the discrete spectrum is indexed by two integers $k$ and $N$, where $k$ is the eigenvalue of the spin-orbit operator, and $N$ can be identified with the winding number of heteroclinic orbits of a certain dynamical system on a compact cylinder that connect a saddle point on one cylinder end to a saddle point on the other.  
This identification allows us to find a one-to-one correspondence between the eigenfunctions of our Hamiltonian and the orbitals of hydrogen found in textbook quantum mechanics that map into the spectrum (\ref{SommerfeldFS}). 
The correspondence allows us to relate $N$ and $k$ to the principal quantum number $n$ by defining $n := N + |k|$. Section 4 of \cite{KLTzGKN} supplies details on this correspondence.
\vspace{-7pt}

\subsection{Outline of the {ensuing sections}}

 {Section \ref{sec: main} is our main technical section. 
After a recap of the results of reference \cite{BMB} in section \ref{sec: state of affairs}, we pave the ground for our approach by introducing suitable dimensionless variables and parameters in section \ref{sec: dimless}, then we invoke the Pr\"ufer transform in section \ref{sec: prufer} to map the reduced Dirac eigenvalue problem into a dynamical system on a half-infinite cylinder.
  In section \ref{sec: conv to dyn syst} we compactify the dynamical system by mapping it to a finite cylinder.
Section \ref{sec: proof} is devoted to the discussion of the compactified dynamical system, establishing a detailed characterization of its orbits, culminating in our main Theorem \ref{thm: main}. 
Section \ref{sec: numerical} features numerically computed eigenvalue branches as functions of $Z$, suggesting that the spectrum of an electron with anomalous magnetic moment in the Reissner--Weyl--Nordstr\"om spacetime is practically indistinguishable from its counterpart in Minkowski spacetime equipped with a point nucleus.
We also emphasize the numerical challenge to resolve the differences in a satisfactory manner. 
We conclude with a summary and outlook in section \ref{sec: sum and out}.}

\section{The Dirac equation of an electron with anomalous magnetic moment in the 
Reissner--Weyl--Nordstr{\"o}m spacetime}\label{sec: main}

\subsection{Summary of the results of ref.\cite{BMB}}\label{sec: state of affairs}

The RWN spacetime, with metric given by \eqref{eq:StaticSO3metric} and \eqref{eq:RWNfSQR}, is a solution to the Einstein--Maxwell(--Maxwell)\footnote{The ``second Maxwell'' refers to his law of the electromagnetic vacuum, while the ``first Maxwell'' stands for the pre-metric Maxwell equations. In our last section we will comment on other electromagnetic vacua, with the ``second Maxwell'' replaced accordingly.} equations with electromagnetic Faraday tensor $F = dA$, where $A$ is the one-form $A = -\frac{Q}{r}cdt$. 
Therefore, in the conventional $(x^0 =ct, r, \theta, \phi)$ coordinates, the only nonzero component of $A$ is $A_0 = -\frac{Q}{r}$. 
Hence, the only nonzero components of $F$ are $F_{0r} = - F_{r0} = \frac{Q}{r^2}$. 

In an electromagnetic spacetime, the Dirac equation for a point electron with mass $\mEL$, charge $-e$, and anomalous magnetic moment $\mu_a$ is given by  (note that each term acting on $\Psi$ has units of energy)
\begin{equation}\label{eq: Dirac}
\tilde{\g}^\mu(i\hbar c\nabla_\mu -e A_\mu)\Psi + \tfrac{1}{2}\mu_a \tilde{\sigma}^{\mu\nu}F_{\mu\nu}\Psi - \mEL c^2 \Psi \,=\,0.
\end{equation}
The anomalous magnetic moment term, $\frac{1}{2}\mu_a\tilde{\s}^{\mu\nu}F_{\mu\nu}$, is the covariant generalization of the corresponding term in flat spacetime, see section 4.2.3 in \cite{ThallerBOOK}. This is accomplished by setting $\tilde{\sigma}^{\mu\nu} = i \tilde{\gamma}^\mu \tilde{\gamma}^\nu$. 
In the following, we leave $\mu_a$ as a parameter.
Physically, the anomalous magnetic moment is measured in multiples of the Bohr magneton: $\mu_a = - a \frac{e\hbar}{2 \mEL c}$, where $a$ is dimensionless. 
The first perturbative term in QED in flat spacetime gives $a = \frac{1}{2\pi}\alpha_{\mbox{\tiny{S}}} \approx 0.00116$; cf. our discussion in the introduction.

By multiplying the Dirac equation \eqref{eq: Dirac} by $\gamma^0$, we can recast it into a Hamiltonian form 
\begin{equation}\label{eq: Dirac Hamil form}
i \hbar \pd_t \Psi \,=\, H \Psi.
\end{equation}
Due to the static nature of the spacetime, the operator $H$ does not depend on time.

For each time $t$, let $\Si_t$ denote the hypersurface of constant $t$ within the $(t,r,\theta, \phi)$ coordinates. We define an inner product on $\Si_t$ via 
\begin{equation}
( \Psi, \Phi )_{\Sigma_t} \,=\, \int_{\Sigma _t} \ov{\Psi} \tilde{\gamma}^\mu \Phi n_\mu dV_{\Sigma_t}.
\end{equation}
Here $\ov{\Psi}$ is the conjugate bispinor defined as $\ov{\Psi} := \Psi^\dagger \gamma^0$, $n$ is the unit future directed timelike normal to $\Sigma_t$, and $dV_{{\Sigma}_t}$ is the volume form on $\Sigma_t$ induced from its Riemannian metric. This choice of inner product can be motivated from an action principle. It can also be motivated from the conservation of current density, i.e. $\nabla_\mu j^\mu = 0$ where $j^\mu = \ov{\Psi}\tilde{\gamma}^\mu\Psi$.

Thus, for each $t$, we have a Hilbert space 
\begin{equation}
\mathcal{H}_t \,=\,\{\Psi \colon M \to \C^4 \, \mid \,(\Psi,\Psi)_{\Sigma_t} <\infty\}.
\end{equation}
The operator $H$ in \eqref{eq: Dirac Hamil form} on the Hilbert space $\mathcal{H}_t$ is called the \emph{Dirac Hamiltonian} on $\Sigma_t$.

The spherical symmetry present within the Lorentzian metric and the electromagnetic field allows one to perform a separation of variables and decompose the Hilbert space $\mc{H}_t$ into a direct sum of partial wave subspaces \cite{CohenPowers}, see also \cite[Sec. 4.6]{ThallerBOOK}.   
The partial wave subspaces are indexed by a triplet $(j, m_j, k_j)$ where 
\begin{equation}
j \,=\, \frac{1}{2}, \frac{3}{2}, \frac{5}{2} \dotsc; \quad m_j \,=\, -j, -j+1, \dotsc, j-1, j; \quad k_j \,=\, \pm \left(j + \half\right).
\end{equation}
These correspond to certain eigenvalues of well-known operators on $L^2(\mathbb{S}^2)^4$, see \cite[p. 126]{ThallerBOOK}. 
The Dirac Hamiltonian $H$ acting on each partial wave subspace is given by a two-dimensional \emph{reduced Hamiltonian} (compare with \cite[eq. (5.46)]{ThallerBOOK}):
\begin{equation}\label{eq: red Hamil}
H_{k_j}\,=\, \begin{pmatrix}
\mEL c^2 f(r) - \frac{eQ}{r} & -\hbar c f^2(r) \pd_r + \hbar c f(r) \frac{k_j}{r} + \mu_af(r)\frac{Q}{r^2}
\\
\hbar c f^2(r) \pd_r + \hbar c f(r)\frac{k_j}{r} +\mu_a f(r) \frac{Q}{r^2} & -\mEL c^2 f(r) - \frac{eQ}{r} 
\end{pmatrix}
\end{equation}
which acts on the Hilbert space
\begin{equation}\label{eq: Hilbert space}
L^2\big((0,\infty), f(r)^{-2}dr\big)^2.
\end{equation}
For a derivation of this form of the reduced Hamiltonian, see \cite[p. 76]{CohenPowers} (see  also \cite{Moulik_Thesis}). 
 Although each partial wave subspace depends on the triplet  $(j,m_j, k_j)$, the reduced Hamiltonian $H_{k_j}$ only depends on the integer $k_j$. 
(The independence of the reduced Hamiltonian on $m_j$ is the reason for the degeneracy of the different spin states within the hydrogen atom; it is a consequence of spherical symmetry.)

The Dirac Hamiltonian $H$ on $\Sigma_t$ is essentially self-adjoint on the domain of $C^\infty$ bispinors compactly supported away from the origin $r = 0$ if and only if each reduced Hamiltonian $H_{k_j}$ is essentially self-adjoint on $C^\infty_c(0,\infty)^2$, in which case, the spectrum of the corresponding self-adjoint operator (still denoted by $H$) is the union of the spectra of the self-adjoint operators for the reduced Hamiltonians (still denoted by $H_{m_j, k_j}$). 
For a proof, see \cite[Lem. 4.15]{ThallerBOOK} which generalizes to our setting. 
This reduces the problem to finding the spectrum for each reduced Hamiltonian.

The following theorem was proved in \cite{BMB}. 
It says that essential self-adjointness is guaranteed for each reduced Hamiltonian $H_{k_j}$ provided the anomalous magnetic moment is not too small.
\medskip

\begin{thm}[Belgiorno--Martellini--Baldicchi\cite{BMB}]\label{thm: Belgiornoa et al}\:\\
Each reduced Hamiltonian $H_{k_j}$ is essentially self-adjoint on $C^\infty_c(0,\infty)^2$ if and only if $|\mu_a| \geq \frac{3}{2}\frac{\sqrt{G}\hbar}{c}$.
In which case, the following hold for the self-adjoint Dirac Hamiltonian $H$:
\begin{itemize}
\item[\emph{(a)}] The essential spectrum of $H$ is $(-\infty, -\mEL c^2] \cup [\mEL c^2, \infty)$;
\item[\emph{(b)}] the purely absolutely continuous spectrum of $H$ is $(-\infty, -\mEL c^2) \cup (\mEL c^2, \infty)$;
\item[\emph{(c)}] the singular continuous spectrum of $H$ is empty;
\item[\emph{(d)}] $H$ has infinitely many eigenvalues in the gap $(-\mEL c^2, \mEL c^2)$ of its essential spectrum. 
\end{itemize}
\end{thm}

\noindent
\begin{rem}
{Recall that the empirical anomalous magnetic moment is $\mu_a^{} = - a \frac{e\hbar}{2 \mEL c}$, where $a$ is dimensionless, see \cite{FanETal}. 
In first perturbative order, QED in flat spacetime gives $a = \frac{1}{2\pi}\alpha_{\mbox{\tiny{S}}} \approx 0.00116$. 
With this \emph{classical value} for the anomalous magnetic moment of the electron, the hurdle for essential self-adjointness is overwhelmingly cleared:
\[
 \hspace{1truecm}
\frac{3}{2}\frac{\sqrt{G}\hbar}{c}\frac{1}{\mu_{\rm cl}} \,=\, 6\pi\frac{\sqrt{G}\mEL}{\alphaS e} \,\approx\, 1.27\cdot 10^{-18} \,\ll\, 1. \hspace{1truecm} \square
\]
}
\end{rem}

Theorem \ref{thm: Belgiornoa et al}(d) implies that the discrete spectrum of the Dirac Hamiltonian on the RWN spacetime has infinitely many eigenvalues. 
 In this paper, we {rigorously investigate the discrete spectrum and classify it.
We also wish to compare} it to the discrete spectrum of the corresponding special relativistic hydrogenic ion problem. 
A naturally tempting conjecture is that the discrete spectrum of the Dirac equation for an electron with anomalous magnetic moment in the RWN spacetime will converge to the discrete spectrum of the corresponding special-relativistic problem in the limit when Newton's $G\to0$, given realistic values for the electron parameters. While we will not attempt to prove this conjecture in the present paper, we have carried out a numerical study of the eigenvalue problem which supports this conjecture.
\medskip

\subsection{Dimensionless variables and parameters}\label{sec: dimless}

The conventional Gaussian units (or, for that matter, also the modern SI units) come equipped with numerical values that obscure the understanding of the Dirac equation for hydrogenic ions more than they illuminate it.
It is advisable to switch to dimensionless quantities by choosing reference units that are more suitable to the atomic realm. 
Items 1 and 2 concern the RWN metric, items 3--4 concern Dirac's equation.
 
\begin{enumerate}

\item The replacement $r \mapsto \frac{\hbar}{\mEL c} r$ renders $r$ dimensionless and measures distance as multiple of the electron's Compton wavelength. 
As a result, $f(r)$ now reads
\begin{equation}\label{eq:fOFrDIMLESS}
f(r) \,=\, \sqrt{1 - 2\frac{GM\mEL}{\hbar c}\,\frac{1}{r} + \frac{G\mEL^2Q^2}{(\hbar c)^2}\,\frac{1}{r^2}}.
\end{equation}

\item 
The coefficients of $\frac1r$ and of $\frac{1}{r^2}$ in (\ref{eq:fOFrDIMLESS}) are now themselves dimensionless parameters, and writing $f(r)$ simply as
\begin{equation}\label{eq: f}
f(r) \,=\, \sqrt{1 - \frac{2A_*}{r} + \frac{Z_*^2}{r^2}}\,
\end{equation}
defines $A_*$ and $Z_*$; note that the condition for a naked singularity becomes $A_* < Z_*$.

 The asymptotic behavior of $1/f^2$ will be useful, so we record it here:
\begin{equation}\label{eq: asympt f}  
\frac{1}{f^2(r)} \,=\, \left\{
\begin{array}{ll}
      Z_*^{-2}r^2 + O(r^3) & \text{ as } r \to 0 \\
      &\:
      \\
      1 + O(r^{-1}) & \text{ as } r \to \infty. 
\end{array} 
\right. 
\end{equation}
\begin{rem}
The motivation for our starred letters $A_*$ and $Z_*$ becomes obvious when comparing (\ref{eq:fOFrDIMLESS}) with (\ref{eq: f}), by recalling that $Q=Ze$ and $M=:A(Z,N)\mPR$, where $\mPR$ denotes the proton mass and $A(Z,N)$ the nuclear mass number.
This reveals that
\begin{align}
A_* \,&:=\, \mc{G}\alphaS\frac{\mPR}{\mEL}A, \label{eq:A_*}
\\
Z_*\,&:=\, \sqrt{\mc{G}}\alphaS Z, \label{eq:Z_*}
\end{align}
where we have also introduced the dimensionless gravitational constant 
\begin{equation}\label{eq: A}
    \mc{G}:=G\mEL^2/e^2.
\end{equation}
\end{rem}

\begin{rem}
What we denote as $A_*$ has been called the gravitational fine structure constant for the interaction of a particle of mass $M$ with one of mass $\mEL$, see \cite{Jentschura}, but we prefer to reserve the name gravitational fine structure constant for that quantity when $M=\mEL$. $\square$
\end{rem}

\item We relabel $k_j$ as $k$ and define the operator $K_k :=\frac{1}{\mEL c^2}H_{k_j}$. 

\item 
In order to facilitate the comparison of our results with those in \cite{ThallerBOOK}, we also introduce $\gamma := -Z\frac{e^2}{\hbar c} = -Z\alpha_{\mbox{\tiny{S}}}$.
Note that $Z_* = -\sqrt{\mc{G}} \g$. 

\item Recalling that $\mu_a = -a\frac{e\hbar}{2\mEL c}$, with $a = \frac{\alphaS}{2\pi}$ to first order in perturbative flat spacetime QED, where $\alphaS$ is Sommerfeld's fine structure constant, we now define $\lambda :=  -\half a \gamma$. 
With our choice of $a$  we have $\lambda = \frac{1}{4\pi}Z\alphaS^2$.

\item For $E \in (-\mEL c^2, \mEL c^2)$, define $\e := \frac{E}{\mEL c^2}$. 
Hence $\e \in(-1,1)$ is dimensionless. In some of the following proofs, $\e = \pm 1$ will be considered even though it is not part of the discrete spectrum by Theorem \ref{thm: Belgiornoa et al}.

\end{enumerate}

\noindent With this new notation, the operator $K_k$ becomes
\begin{equation}\label{eq: red Hamil K}
K_k \,=\, \begin{pmatrix}
f(r)+\frac{\gamma}{r} &  -f^2(r) \pd_{r} +  f(r) \frac{k}{r} - f(r) \frac{\lambda}{ r^2}
\\
f^2(r)\pd_{ r} +  f(r)\frac{k}{ r} - f(r) \frac{\lambda}{ r^2} &-f(r)+\frac{\gamma}{ r} 
\end{pmatrix}.
\end{equation}

\begin{rem}
{Note that $f(r)$, given in (\ref{eq: f}), is the only place in $K_k$ where Newton's gravitational constant $G$, or for that matter its dimensionless version $\mc{G}$, enters, through $A_*$ and $Z_*$.
Formally, the $\mc{G} \to 0$ limit yields $f(r)\to 1$ for $r>0$, and the reduced Hamiltonian \eqref{eq: red Hamil K} defined on its minimal domain $C^\infty_c(0,\infty)^2$ becomes the reduced Hamiltonian in the corresponding special-relativistic problem, see \cite[eq. (7.167))]{ThallerBOOK}. $\square$
}
\end{rem}

\subsection{The Pr{\"u}fer transformation}\label{sec: prufer}

The eigenvalue problem for the reduced Hamiltonian yields a coupled system of differential equations. 
 In this section, we use a Pr\"ufer transformation to an equivalent system of
two differential equations where one of them is decoupled from the other. 
This technique has been recognized in the past to help numerically solve eigenvalue problems for Dirac operators \cite{ULEHLA1986355}.

Note that $E$ is an eigenvalue for $H_{m_j,k_j}$ if and only if $\e$ is an eigenvalue for $K_k$, where $K_k$ is given by \eqref{eq: red Hamil K}. 
The eigenvalue problem 
\begin{equation}
K_k \begin{pmatrix} u \\ v \end{pmatrix}
\,=\, 
\e \begin{pmatrix} u \\ v \end{pmatrix}
\end{equation}
yields the following coupled system of first order linear differential equations
\begin{align}
&u'(r) + \frac{1}{f}\left(\frac{k}{ r} - \frac{\l}{r^2}\right)u(r) - \frac{1}{f^2}\left(f-\frac{\g}{r} + \e \right)v(r)
 \,=\,0 \label{eq: du/dr}\\
&v'(r)  - \frac{1}{f}\left(\frac{k}{r} - \frac{\l}{r^2}\right)v( r) - \frac{1}{f^2}\left(f + \frac{\gamma}{r} - \e \right)u(r) \,=\, 0. \label{eq: dv/dr}
\end{align}

We introduce the variables $R(r)$ and $\Omega(r)$ via a Pr{\"u}fer transform:
\begin{equation}
R(r) \,:=\, \sqrt{u^2(r) + v^2(r)} \quad \text{ and } \quad \Omega(r) \,:=\, 2\arctan\left(\frac{u(r)}{v(r)} \right),
\end{equation}
where $R \in [0, \infty)$ and $\Omega \in (-\pi, \pi]$. 
Then 
\begin{equation}\label{eq: u and v def}
u(r) \,=\ R \cos \left(\half \Omega \right) \quad \text{ and } \quad v(r) \,=\, R \sin \left(\half \Omega \right).
\end{equation}

Summing $\eqref{eq: du/dr} \cdot u$ and $\eqref{eq: dv/dr} \cdot v$ gives
\begin{equation}\label{eq: R'}
\frac{R'}{R} \,=\, \frac{1}{f}\sin \Omega + \frac{1}{f}\left(-\frac{k}{r} + \frac{\l}{r^2} \right) \cos \Omega.
\end{equation}
Likewise, summing $\eqref{eq: dv/dr}\cdot u$ and $-\eqref{eq: du/dr} \cdot v$ gives 
\begin{equation}\label{eq: Omega'}
\Omega' \,=\, \frac{2}{f}\cos \Omega + \frac{2}{f}\left(\frac{k}{r} - \frac{\l}{r^2} \right) \sin \Omega + \frac{2}{f^2}\left(\frac{\g}{r} - \e\right).
\end{equation}
Although \eqref{eq: du/dr} and \eqref{eq: dv/dr} is a linear system and \eqref{eq: R'} and \eqref{eq: Omega'} is nonlinear, the advantage we gain is that the equation for $\Omega'$ does not depend on $R$. Therefore we only need to study the differential equation  \eqref{eq: Omega'} for $\Omega$ and then $R$ can be solved by integrating \eqref{eq: R'}.

Once we solve for $\Omega(r)$ and $R(r)$, we can define $u(r)$ and $v(r)$ via \eqref{eq: u and v def}. If $u$ and $v$ belong to the Hilbert space $L^2\big((0,\infty), f(r)^{-2}dr^2\big)$, which holds if and only if $R/f \in L^2\big((0,\infty), dr\big)$, then the corresponding wave function $\Psi$ represents an element of the Hilbert space $\mc{H}_t.$

\medskip

\subsection{Conversion to a dynamical system on a compact cylinder}\label{sec: conv to dyn syst}

The study of \eqref{eq: Omega'} is facilitated by converting the differential equation into a dynamical system on a compact cylinder. 
First, we make the system autonomous by introducing a new differential equation, $r' = 1$, so that $r$ is now a dependent variable. 
However, the autonomous two-dimensional system, $r' = 1$ and \eqref{eq: Omega'}, is not compact and is singular at the origin $r = 0$. We rectify both of these concerns. 
We introduce transformations that will make the system compact and remove the singularity at the origin $r = 0$.

We first compactify the system by bringing $r = \infty$ into a finite value. 
Introduce a new independent variable $\eta$ via the transformation
\begin{equation}
\eta: = T(r) \,:=\, \frac{r}{1+r} \quad \text{ with inverse } \quad r \,=\, T^{-1}(\eta) = \frac{\eta}{1-\eta}.
\end{equation}
$T$ maps $(0,\infty)$ diffeomorphically onto $(0,1)$. 
From \eqref{eq: f}, we have
\begin{equation}
f\big(r\big) =:\frac{1}{\eta}g(\eta),
\end{equation}
where
\begin{equation}\label{eq: g(eta)}
g(\eta) \,=\, \sqrt{(1 + 2A_* + Z_*^2)\eta^2 -2(A_* + Z_*^2)\eta + Z_*^2}.
\end{equation}
Note that $g(\eta)$ is always positive since we are in the naked singularity sector, $A_* < Z_*$.

Similar to \eqref{eq: asympt f}, we have
\begin{equation}\label{eq: asympt g}  
\frac{1}{f^2\big(r\big)} 
\,=\, \frac{\eta^2}{g^2(\eta)} \,=\, \left\{
\begin{array}{ll}
      \frac{\eta^2}{Z_*^2} + O(\eta^3) & \text{ as } \eta \to 0 \\
      &\:
      \\
      1 + O(\frac{1-\eta}{\eta}) & \text{ as } \eta \to 1. 
\end{array} 
\right. 
\end{equation}
The system, $r' = 1$ and \eqref{eq: Omega'}, becomes
\begin{equation}\label{eq: precpt syst}
\left\{
\begin{array}{ll}
\eta' &=\, (1-\eta)^2 
\\
\Omega' &=\, \frac{2\eta}{g}\cos \Omega + \frac{2}{g}\left(k(1-\eta) - \lambda \frac{(1-\eta)^2}{\eta} \right)\sin \Omega + \frac{2\eta}{g^2}\big(\g(1-\eta)-\e\eta \big). 
\end{array}
\right.
\end{equation}

The system \eqref{eq: precpt syst} is now precompact. 
It is not compact since, unless $\lambda=0$, there is still a singularity at $\eta = 0$ (the coefficient for the anomalous magnetic moment $\lambda$ behaves like $1/\eta$ as $\eta \to 0$). 

\begin{rem}
    The fact that $\eta = 0$ is a regular point of this dynamical system when there is no anomalous magnetic moment term, is connected with the loss of essential self-adjointness for the Dirac Hamiltonian on the RWN background, first noted by \cite{CohenPowers}, that was discussed in the Introduction.
\end{rem}

To rectify this issue with \eqref{eq: precpt syst}, we introduce a new independent variable $\tau$ via $\tau := r + \log(r)$, so that $\frac{dr}{d\tau} = \eta$. 
Multiplying $\eta'$ and $\Omega'$ by $\eta$ in \eqref{eq: precpt syst}  and introducing the notation $\dot{\Omega} = \frac{d\Omega}{d\tau}$ and $\dot{\eta} = \frac{d\eta}{d\tau}$, the system becomes
\begin{equation}\label{eq: cpt syst}
\left\{
\begin{array}{ll}
\dot{\eta} &=\, \eta(1-\eta)^2
\\
\dot{\Omega} &=\, \frac{2\eta^2}{g} \cos \Omega + \frac{2}{g}\big(k\eta(1-\eta) - \lambda (1-\eta)^2 \big)\sin \Omega + \frac{2\eta^2}{g^2}\big(\g (1-\eta) - \e \eta \big). 
\end{array}
\right.
\end{equation}
The system \eqref{eq: cpt syst} is now compact and free of singularities. 
Understanding the solutions to this system will be the main objective for the remainder of the paper.

\begin{rem} 
We note that, in addition to the eigenvalue parameter $\varepsilon \in (-1,1)$, the above dynamical system depends on three other continuous parameters: $\lambda \in \Rset$, representing the non-dimensionalized value of the electron anomalous magnetic moment, $\mathcal{G} \in \Rset_+$, the non-dimensionalized universal constant of gravitation, and $m_p/m_e$, the ratio of proton to electron rest mass (the latter two parameters are hiding inside the function $g$), as well as the two discrete parameters $k\in \Zset$ and $Z \in \Nset$.  
 With so many parameters one should expect the presence of different regions in the parameter space with significantly different qualitative behavior for the system, and a wealth of bifurcation phenomena. For example, setting $\mathcal{G} = 0$, i.e.
in the case of Minkowski spacetime as the background, with $f \equiv 1$ (and hence $g(\eta) = \eta$), the system \eqref{eq: cpt syst} is still not compactified.  
One would have to multiply $\eta'$ and $\Omega'$ in \eqref{eq: precpt syst} by $\eta^2$ instead of $\eta$. 
In this case, the independent variable, call it $\tilde \tau$, would have to satisfy $\frac{dr}{d\tilde \tau} = \eta^2$. This implies that the equilibrium points at $\eta=0$ will be non-hyperbolic, giving rise to qualitatively different local phase portraits from those studied below. As another example, the restriction $\lambda\geq \frac{3}{2}Z_* = \frac{3\alphaS Z}{2}\sqrt{\mathcal{G}}$ that we encountered in Theorem~\ref{thm: Belgiornoa et al} and will see again in Theorem~\ref{thm: L^2} is connected with the presence of a bifurcation curve in the $(\mathcal{G},\lambda)$ plane. $\square$
\end{rem}

The system \eqref{eq: cpt syst} now forms a dynamical system on the compact cylinder $\mc{C} = [0,1] \times \mathbb{S}^1$. 
There are only four equilibrium points. 
If we identify $\mc{C}$ with the fundamental domain $\mc{C}_* = [0,1] \times [-\pi, \pi)$, these four points are:
\begin{equation}\label{eq: equi pts}
S^- \,:=\, (0,0), \quad N^- \,:=\, (0,-\pi), \quad S^+_\e \,:=\, (1, -\arccos\e), \quad N^+_\e \,:=\, (1, \arccos\e).
\end{equation}
Note that $S^-$ and $N^-$ appear on the left boundary of the cylinder, while $S^+_\e$ and $N^+_\e$ appear on the right boundary of the cylinder. 
The Jacobians are
\begin{align*}
J(S^-) \,=\, \begin{pmatrix}
1 & 0
\\
0 & -2\frac{\l}{Z_*}
\end{pmatrix} \quad &\text{ and } \quad J(N^-) \,=\, \begin{pmatrix}
1 & 0
\\
0 & 2\frac{\l}{Z_*}
\end{pmatrix},
\\
J(S^+_\e) \,=\, \begin{pmatrix}
0 & 0
\\
c_+ & \sqrt{1-\e^2}
\end{pmatrix}
\quad &\text{ and } \quad J(N^+_\e) \,=\, \begin{pmatrix}
0 & 0
\\
c_- & -\sqrt{1-\e^2}
\end{pmatrix},
\end{align*}
where
\[
c_{\pm} \,:=\, 2(1-A_* \pm k)\sqrt{1-\e^2} +2\e(2A_*-1) - 2\g.
\]

$S^-$ and $N^-$ are hyperbolic equilibrium points; they correspond to a saddle and node, respectively. 
From the stable manifold theorem \cite{Perko}, there is a unique (up to translation by a constant) orbit emanating out of $S^-$ into the cylinder, called the \emph{unstable manifold}. 
$S^+_\e$ and $N^+_\e$ are non-hyperbolic equilibrium points, and so the stable manifold theorem does not apply. 
They correspond to saddle-nodes with the saddle part of $S^+_\e$ pointing into the cylinder, and the saddle part of $N^+_\e$ pointing out of the cylinder; see Theorem 1 in section 2.11 of \cite{Perko}. 
From Theorem 2.19 in \cite{QTPDS}, it follows that there is a unique (up to translation by a constant) orbit flowing into $S^+_\e$ which will be referred to as its \emph{stable manifold}.

A \emph{heteroclinic orbit} is an orbit of the dynamical system whose $\a$ and $\omega$-limit sets are equilibrium points. 
For the system \eqref{eq: cpt syst}, a heteroclinic orbit is one that begins and ends at the equilibrium points. Since $\dot{\eta} > 0$ for $\eta \in (0,1)$, a heteroclinic orbit that does not lie on the boundary, $\eta = 0$ or $\eta = 1$, must connect one of the equilibrium points $S^-$ or $N^-$ to either $S^+_\e$ or $N^+_\e$; these will be referred to as \emph{non-boundary heteroclinic orbits.} 
A non-boundary heteroclinic orbit that begins at $S^-$ most likely ends at $N^+_\e$. If, in the rare situation, a non-boundary heteroclinic orbit joins $S^-$ to $S^+_\e$, then the orbit will be called a \emph{saddles connector.} 

\medskip

\begin{thm}\label{thm: L^2}
Assume $\l \geq \frac{3}{2}Z_*$.
Consider a non-boundary heteroclinic orbit of the system \eqref{eq: cpt syst}. 
Then this orbit represents an element of the Hilbert space, via  \eqref{eq: u and v def}, if and only if the orbit is a saddles connector.
\end{thm}

\proof
As shown below \eqref{eq: Omega'}, this orbit is an element of the Hilbert space if and only if $R/f \in L^2\big((0,\infty), dr\big)$. 
From \eqref{eq: R'}, we have
\[
R(r)\,=\, \exp\left(\int_{r_0}^r\frac{\sin \Omega(\tilde{r})}{f(\tilde{r})} d\tilde{r}  \right)\exp\left(\int_{r_0}^r \frac{\cos \Omega(\tilde{r})}{f(\tilde{r})}\left(-\frac{k}{\tilde{r}} + \frac{\lambda}{\tilde{r}^2} \right) d\tilde{r}\right)
\]
for some $r_0 \in (0, \infty)$. 
We first examine the $L^2$ condition for the left end of the cylinder, $\eta = 0$ (i.e. $r = 0$). 
Using \eqref{eq: asympt f} in the above expression for $R(r)$, we conclude that 
\begin{itemize}
    \item[(i)] If the orbit begins at $S^-$, then $R(r) \propto r^{\l/Z_*} + o(1)$ as $r \to 0$. 
    \item[(ii)] If the orbit begins at $N^-$, then $R(r) \propto r^{-\l/Z_*} + o(1)$ as $r \to 0$. 
\end{itemize}
Therefore $R^2(r)/f^2(r) \propto r^{2\pm 2\l/Z_*} + o(1)$ as $r \to 0$. 
The $L^2$ condition is satisfied in a neighborhood of $r = 0$ if and only if $2 \pm 2\l/Z_* > -1$. 
Therefore (i) implies the $L^2$ condition in a neighborhood of $r = 0$ while (ii) does not so long as $\l \geq \frac{3}{2}Z_*$.

Now we examine the $L^2$ condition for the right end of the cylinder, $\eta = 1$ (i.e. $r = \infty$). 
Fix $\a > 0$ such that $\a < \sqrt{1-\e^2}$. 
Set $\beta_{\pm} := \sqrt{1-\e^2}  \pm \a > 0$. 
If the orbit ends at $S^+_\e$, then $\sin \Omega/f = -\sqrt{1-\e^2} + o(1)$ as $r \to \infty$. 
Therefore there is an $R_\a$ such that $\sin\Omega(r)/f(r)<-\beta_-$ whenever $r>R_\a$. 
Using similar estimates for $\cos \Omega/f$, we observe:
\begin{itemize}
    \item[(iii)] If the orbit ends at $S^+_\e$, then for some constants $c_1$ and $c_2$, we have the following:
    \begin{itemize}
        \item [$\bullet$] If $k > 0$, then
        $$
         R(r) \leq c_1e^{-r\beta_-}\frac{e^{-\l\beta_+/r}}{r^{k\beta_-}}
        \leq c_1e^{-r\beta_-}\frac{1}{r^{k\beta_-}} 
        \leq c_1 e^{-r\beta_-} $$
        for all sufficiently large $r$.
        \item[$\bullet$] If $k < 0$, 
        then
       $$
         R(r) \leq c_2e^{-r\beta_-}\frac{e^{-\l\beta_+/r}}{r^{k\beta_+}}
         \leq c_2e^{-r\beta_-}\frac{1}{r^{k\beta_+}} 
       \leq c_2 e^{-r\beta_-/2} 
       $$
       for all sufficiently large $r$.
    \end{itemize}
In either case, the $L^2$ condition is satisfied for $R^2/f^2$ in a neighborhood of $r = \infty$. 
\end{itemize}
\begin{itemize}
    \item[(iv)] If the orbit ends at $N^+_\e$, then $\sin \Omega/f = \sqrt{1-\e^2} + o(1)$ as $r \to \infty$. Hence $R(r)$ exhibits exponential growth as $r \to \infty$ and so the $L^2$ condition is never satisfied. \qed
\end{itemize}

\begin{rem}
{
If we instead considered negative values of $\l$ in Theorem \ref{thm: L^2}, then $N^-$ and $S^-$ would have switched roles and we would have required $|\lambda| \geq \frac{3}{2}Z_*$.
The assumption $|\lambda| \geq \frac{3}{2}Z_*$ is equivalent to the assumption $|\mu_a| \geq \frac{3}{2}\frac{\sqrt{G}\hbar}{c}$, which appeared in Theorem \ref{thm: Belgiornoa et al}. 
In fact Theorem \ref{thm: L^2} offers another proof of the fact that $H$ is essentially self-adjoint if and only if $|\mu_a| \geq \frac{3}{2}\frac{\sqrt{G}\hbar}{c}$, see \cite[Thm. 4.16]{ThallerBOOK}. $\square$
 }
\end{rem}

\subsection{Proof of the main result}\label{sec: proof}

In this section, we will always assume the following conditions on our parameters
\[
Z_* \,>\,0, \quad \quad 0 \leq A_* < Z_*\, \quad \quad \lambda \geq \frac{3}{2}Z_*.
\]
Recall that the second assumption implies $g(\eta) > 0$ (naked singularity sector), and the last assumption guarantees essential self adjointness. 
Also, the other parameter is $k$ which is a nonzero integer. 
Lastly, recall that $\gamma$ and $Z_*$ are not independent parameters but are related via $Z_* = -\sqrt{\mc{G}}\gamma$ where $\mc{G}$ is given by \eqref{eq: A}.

Given Theorem \ref{thm: L^2}, we wish to obtain saddles connectors for our system, which is our main result, Theorem \ref{thm: main}, at the end of this section. 
We will follow the strategy first laid out in \cite{KieTah14a} and generalized in \cite{KLTzGKN}; the latter will be our primary reference.  Our compact dynamical system on the cylinder $\mc{C} = [0,1] \times \mathbb{S}^1$ is given by 
\begin{equation}
\left\{
\begin{array}{ll}
\dot{\eta} &=\, \mathfrak f(\eta)
\\
\dot{\Omega} &=\, \mathfrak{g}_\e(\eta, \Omega). 
\end{array}
\right.
\end{equation}
where $\mathfrak{f}(\eta)$ and $\mathfrak{g}_\e(\eta, \Omega)$ are given by the right hand sides of \eqref{eq: cpt syst}.

So long as $-1 < \e <1$, the dynamical system $\mc{C}$ satisfies properties (a) - (e) in section 3.1 of \cite{KLTzGKN} with $x = \eta$, $y = \Omega$, parameter $\mu = \e$, and fundamental domain  $\mc{C}_* = [0,1] \times [-\pi, \pi)$.

We briefly state these properties in our context.

\begin{enumerate}

\item[\bf{(a)}] $F(0) = F(1) = 0$ and $\mathfrak{f}(\eta) > 0$ for all $\eta \in (0, 1)$. Consequently, equilibrium points can only occur at the boundary of the cylinder (i.e. at $\eta = 0,1$), and the boundaries $\eta = 0$ and $\eta = 1$ are images of orbits of the dynamical system. 
An orbit whose image is not completely contained in one of the boundaries $\eta = 0$ or $\eta = 1$ will be referred to as a \emph{non-boundary orbit}. Since $\mathfrak{f}(\eta) > 0$ for $\eta \in (0, 1)$, the flow of the dynamical system ``points to the right." 
That is, for any non-boundary orbit $\big(\eta(\tau), \Omega(\tau)\big)$, we have $\eta(\tau)$ is strictly increasing in $\tau$.

\item[\bf{(b)}] $S^-$ and $N^-$, from \eqref{eq: equi pts}, are hyperbolic equilibrium points; they correspond to a saddle and node, respectively. $S^+_\e$ and $N^+_\e$ are non-hyperbolic equilibrium points. We will let $\mc{W}^-_\e$ and $\mc{W}^+_\e$ denote the unstable and stable manifolds associated with $S^-$ and $S^+_\e$, respectively. Note that $\mc{W}^-_\e$ and $\mc{W}^+_\e$ coincide coincide when the orbit is a saddles connector.

\item[\bf{(c)}] Set $\mc{C}_* = [0, 1] \times [-\pi,\pi)$ to be a fundamental domain of the universal cover $\wt{\mc{C}} = [0, 1] \times \RR$. 
The equilibrium points $N^+_\e$ and $S^+_\e$ stay within this fundamental domain for all $\e \in (-1,1)$ and $N^+_\e$ is always `above' $S^+_\e$, that is,  $-\pi \leq -\arccos \e < \arccos \e < \pi.$

\item[\bf{(d)}]  For all $(\eta,\Omega) \in \mc{C}$, we have $\frac{\pd}{\pd \e}\mathfrak{g}_\e(\eta,\Omega) \leq 0$. 

\item[\bf{(e)}] Every non-boundary orbit is a heteroclinic orbit. 
That is, the $\a$ and $\omega$-limit sets of any non-boundary orbit are precisely one-point sets containing equilibrium points on the boundary $\eta = 0$ and $\eta= 1$, respectively.  
Therefore we will refer to the $\alpha$ and $\omega$-limit sets as $\a$ and $\omega$-limit \emph{points}.

\end{enumerate}

Only property (e) requires some justification. 
A quick calculation shows that Proposition 3.16 from \cite{KLTzGKN} holds for $\e \in (-1,1]$. (Note that $\e$ is analogous to $E$ in \cite{KLTzGKN}.) 
A similar argument as used in Lemma 3.17 from \cite{KLTzGKN} also holds. 
This combined with Theorem 3.5 in \cite{KLTzGKN} shows that assumption (e) holds.

Therefore the consequences of properties (a) - (e) laid out in section 3.2 of \cite{KLTzGKN} hold for our system which we briefly discuss. 
Associated to the dynamical system on $\mc{C}$ is a topological number called the \emph{winding number} $w_\e$ \cite[Def. 3.2]{KLTzGKN} which counts the number of times the unstable manifold $\mc{W}^-_\e$ winds around the cylinder. 
Specifically, let $\wt{\mc{W}}^-_\e$ denote the lift of $\mc{W}^-_\e$ to the universal cover $\wt{C} = [0,1] \times \RR$ beginning at $(0,0)$. 
Then $\wt{\mc{W}}^-_\e$ terminates at either a lift of the nodal point $N^+_\e$ or a lift of the saddle point $S^+_\e$; hence it terminates at $(1, -2\pi N \pm \arccos\e)$ for some $N \in \Z$. 
This number $N$ is precisely the winding number.

To establish existence and uniqueness of saddles connectors with different winding numbers, we rely on the following proposition. 

\medskip

\begin{prop}\label{prop: green}
Let $N$ be an integer. 
Suppose there exist two parameter values $\e' < \e''$ such that 
\[
w_{\e'} \,\leq\, N \quad \text{ and } \quad w_{\e''}\,\geq\, N+1.
\]
Then there exists a unique $\e \in [\e', \e'')$ such that $\mc{W}^-_\e$ is a saddles connector with winding number $w_\e = N$.
\end{prop}

\proof
Existence of the saddles connector follows from Proposition 3.4 in \cite{KLTzGKN}. 
To prove uniqueness, suppose there were two values $\e_1 < \e_2$ which correspond to saddles connectors $\mc{W}^-_1$ and $\mc{W}^-_2$, respectively, each with the same winding number $N$. 
Let $\wt{\mc{W}}^-_1$ and $\wt{\mc{W}}^-_2$ denote their lifts into the universal cover $\wt{\mc{C}} = [0,1] \times \RR$ whose $\a$-limit point is (0,0). 
By Lemma 3.3 in \cite{KLTzGKN}, it follows that $\wt{\mc{W}}^-_2$ lies below $\wt{\mc{W}}^-_1$. 
However, the $\omega$-limit points of $\wt{\mc{W}}^-_1$ and $\wt{\mc{W}}^-_2$ occur at $\Omega_1 = -2\pi N-\arccos\e_1$ and $\Omega_2 =  - 2\pi N - \arccos\e_2$, respectively. 
Since $\Omega_2 < \Omega_1$, it follows that $\wt{\mc{W}}_2^-$ lies above $\wt{\mc{W}}_1^-$ sufficiently close to the right hand side of the boundary ($\eta= 1$). 
This is a contradiction.
\qed

\medskip

Proposition \ref{prop: green} suggests a strategy of using barriers to constrain the winding numbers. 
For example, the proof of the next proposition shows that negative winding numbers are impossible since the slope field is always negative along the horizontal line $\Omega = \pi$ (in the universal cover).

\medskip

\begin{prop}\label{prop: upper barrier}
For any $\e\in(-1,1)$, the winding number satisfies $w_\e \geq 0$.
\end{prop}

\proof
We work in the universal cover $\wt{\mc{C}}$. 
From \eqref{eq: cpt syst}, the slope field at $\Omega = \pi$ is
\[
\dot{\Omega}|_{\Omega = \pi}(\eta) \,=\, -\frac{2\eta^2}{g(\eta)} + \frac{2\eta^2}{g^2(\eta)}\big(\g(1-\eta) -\e \eta). 
\]
We want to show the above expression is negative for all $\eta \in (0,1)$. 
It is sufficient to show it is negative for $\e = -1$. 
Therefore it suffices to show $-g(\eta) + \g(1-\eta) + \eta < 0$. Since $g$ is positive, this expression is clearly negative for $\eta \in (0, \eta_0)$ where $\eta_0 = -\g/(1-\g)$. 
For $\eta \in [\eta_0, 1)$, it suffices to show $\big[\g(1-\eta) + \eta\big]^2 < g^2(\eta)$. 
Collecting terms in $\eta$, this is equivalent to 
\[
\eta^2\big[(\g^2 - 2\g) - (2A_* + Z_*^2) \big] + \eta\big[2\g(1-\g) + 2(A_* + Z_*^2) \big] - Z_*^2 \,<\, 0.
 \]
Assume for now that the factor multiplying $\eta^2$ is positive. 
Then this quadratic in $\eta$ is concave up. At $\eta = 0$, the quadratic evaluates to $-Z_*^2 < 0$. 
At $\eta = 1$, it evaluates to $-\g^2 <0$. 
Therefore the quadratic is negative between $0$ and $1.$

It suffices to show $\g^2 - 2\g > 2A_* + Z_*^2$. Recall $\g = -Z_*/\sqrt{\mc{G}}$ where $\mc{G}$ is given by \eqref{eq: A}. 
Hence it suffices to show $A_* < Z_*^2 (1-\mc{G})/2\mc{G} + Z_*/\sqrt{\mc{G}}$. 
Indeed this inequality follows from the naked singularity sector hypothesis, $A_* < Z_*$.
\qed

\medskip

An open problem in \cite{KieTah14a} was to show the existence of winding numbers for all non-negative integers.
This problem was resolved in \cite{KLTzGKN} where the key argument is Proposition 3.22 in that paper. 
An analogue of that proposition holds for our system \eqref{eq: cpt syst} as well.

\medskip

\begin{prop}\label{prop: large wind num}
For any integer $N\geq 0$, there is an $\e'' \in (-1,1)$ such that $w_{\e''} \geq N + 1$.
\end{prop}

\proof
The proof is by contradiction and follows the same strategy as the proof of Proposition 3.22 in \cite{KLTzGKN}. 
Suppose there is an integer $N \geq 0$ such that for all $\e \in (-1,1)$, the winding number $w_\e$ is no greater than $N$. 
By Proposition \ref{prop: upper barrier}, the lift of the unstable manifold $\wt{\mc{W}}^-_\e$ emanating from $S^-$ is contained in the compact set $[0,1] \times [-\pi - 2\pi N, \pi]$ for all $\e$. 
It follows that, as $\e \to 1$, the unstable manifolds $\wt{\mc{W}}^-_\e$ converge to an orbit of the dynamical system; see Proposition 3.24 in \cite{KLTzGKN}. 
Call this limiting orbit $c(\tau)$. 
Proposition 3.18 in \cite{KLTzGKN} holds for our system \eqref{eq: cpt syst} as well. 
This fact together with the Poincar{\'e}--Bendixson theorem implies that $c(\tau)$ must converge to an equilibrium point on the boundary $\eta = 1$. 
For $\e = 1$, the equilibrium points $S^+_1$ and $N^+_1$ are both equal to the point $(1,0)$ in the $(\eta, \Omega)$ plane. 
Therefore $c(\tau)$ must converge to some copy of this point in the universal cover.

However at $\e = 1$, the equilibrium point $(1,0)$ is a nilpotent singularity. 
We want to apply Theorem 3.5 in \cite{QTPDS} to \eqref{eq: cpt syst}. 
To do this, we shift our system $\eta_1 = \eta - \frac{\pi}{2}$ so that the equilibrium point is at the origin $(0,0)$ in the $(\eta_1, \Omega)$ plane. 
We also perform the compression $\eta_2 = -2\gamma \eta_1$. 
Our system becomes
\[
\left\{
\begin{array}{ll}
\dot{\eta}_2 &=\, -\frac{1}{2\g}\eta^2_2\left(1 -\frac{\eta_2}{2\gamma} \right) 
\\
\dot{\Omega} &=\, \eta_2 + A(\eta_2, \Omega).
\end{array}
\right.
\]
$A(\eta_2,\Omega)$ contains quadratic terms and higher. 
(The point of the compression was to ensure that the constant in front of $\eta_2$ in the expression for $\dot{\Omega}$ equals 1, so we can directly apply Theorem 3.5 in \cite{QTPDS}.)  
Let $\eta_2 = f(\Omega)$ be the solution to $\eta_2 + A(\eta_2, \Omega) = 0$. 
Using the implicit function theorem, one calculates:  $f'(0) = 0$ and $f''(0) = 2$. 
Therefore $f(\Omega) = \Omega^2 + \text{h.o.t.}$ 
Define $F(\Omega) = B\big(f(\Omega)\big)$ where $B = \dot{\eta}_2$. 
Then $F(\Omega) = -\frac{1}{2\g}\Omega^4 + \text{h.o.t.}$ Define
 \[
 G(\Omega) \,=\, \left(\frac{\pd A}{\pd \Omega} + \frac{\pd B}{\pd \eta_2}\right)\bigg|_{\big(f(\Omega), \Omega\big)} \,=\, -2\Omega + \text{h.o.t.}
 \]
Applying Theorem 3.5(4) in \cite{QTPDS} with $x = \Omega$ and $y =\eta_2$ shows that $m = 4$ and $n =1$; hence $(i2)$ in that theorem is the relevant scenario for our system. 
The phase portrait depicted in Figure 3.21(i) in \cite{QTPDS} shows that no orbit starting within the cylinder (i.e. a non-boundary orbit) can reach the equilibrium point. 
\qed

\medskip

Recall that $\gamma$ and $Z_*$ are not independent parameters but are related via $Z_* = -\sqrt{\mc{G}}\gamma$ where $\mc{G}$ is given by \eqref{eq: A}.  
An assumption on $\gamma$ implicitly defines an assumption on $Z_*$. 
For example, $\g \in [-1,0)$ is equivalent to $Z_* \in (0,\sqrt{\mc{G}}\,]$; this is relevant for the next proposition.

\medskip

\begin{prop}\label{prop: lower barrier for neg k}
For $\e = 0$, the winding number satisfies $w_{0} = 0$ for all $\g \in [-1, 0)$, $A_* \in [0, \half Z_*]$, $\l \geq \frac{3}{2}Z_*$, and $k\leq -1$.
\end{prop}

\proof
By Proposition \ref{prop: upper barrier}, it suffices to show $w_{0} \leq 0$.
We show that $\Omega = -\frac{\pi}{2}$ is a lower barrier. 
That is, we will establish the following inequality: 
\[
\dot{\Omega}|_{\Omega = -\frac{\pi}{2},\,\e = 0}(\eta) \,=\, \frac{2}{g(\eta)}\big(- k\eta(1-\eta) + \lambda(1-\eta)^2\big) + \frac{2\eta^2}{g^2(\eta)}\g(1-\eta)\,>\,0.
\]
Eliminating the positive factor $(1-\eta)$ and rationalizing the denominator, we want to show
\[
g^2(\eta)\big( -k\eta + \lambda(1-\eta) \big)^2 - \g^2\eta^4 \,>\,0 \quad\quad \text{ for all } \quad\quad \eta \in (0,1).
\]
The above expression decreases as $k$ increases, so it suffices to show that the inequality holds for $k = -1$. 
Similarly, it suffices to show that it holds for $\lambda = \frac{3}{2}Z_*$ and $A_* = \half Z_*$.  
Therefore the desired inequality is
\[
p(\eta) \,>\,0,
\]
where 
\[
p(\eta) \,:=\, g^2(\eta)\big|_{A_* = \frac{Z_*}{2}}\left(\eta + \frac{3Z_*}{2}(1-\eta) \right)^2 - \g^2\eta^4.
\]
We define
\[
p_0(\eta) \,:=\, p(\eta)\big|_{Z_* = 0} \,=\, \eta^4(1-\g^2)\quad \text{ and } \quad p_1(\eta) \,:=\, p(\eta) - p_0(\eta).
\]
Since $\g \in [-1,0)$, we have $p_0(\eta) \geq 0$ for $\eta \in [0,1]$. 
Therefore it remains to show that $p_1(\eta) > 0$ for $\eta \in (0,1)$. 
We find
\[
p_1(\eta) \,=\,\frac{Z_*(1-\eta)}{4}\big( a_3\eta^3 + a_2\eta^2 + a_1\eta + a_0\big),
\]
where
$$
a_3 = -9Z_*^3 + 3Z_*^2 -Z_* + 8,
\quad
a_2 = Z_*(27Z_*^2 - 6Z_* + 1),
\quad
a_1 = 3Z_*^2(-9Z_* + 1),
\quad
a_0 = 9Z_*^3.
$$
Recall that $\g \in [-1,0)$ is equivalent to $Z_* \in (0,\sqrt{\mc{G}}\,]$. 
Since $\sqrt{\mc{G}} \approx 4.90\cdot 10^{-22}$, it is not hard to estimate the maximum and minimum values of the coefficients $a_i$ as $Z_*$ runs through $(0,\sqrt{\mc{G}}\,]$. 
Indeed, it follows that for all  $Z_* \in (0,\sqrt{\mc{G}}\,]$, the coefficients $(a_3,a_2,a_1,a_0)$ have signs $(+,+,+,+)$. 
Therefore the result follows by Descartes' rule of signs. \qed

\medskip

\begin{rem} 
The condition $A_* \in [0, \half Z_*]$ is not strict and can be improved. 
Ideally, one would want Proposition \ref{prop: lower barrier for neg k} to hold for all $A_* \in [0,Z_*)$. 
However,  $p_1(\eta)$ is negative near $Z_*$ in the case  $A_* = Z_*$. 
Physically, if $Z_*$ represents the charge of the proton (i.e. $Q = e$), then $A_* = \half Z_*$ is on the order of $10^{-7}$ grams which is about $10^{17}$ proton masses. $\square$
\end{rem}

\medskip

\begin{prop}\label{prop: saddles connector for negative k}
Suppose $\g \in [-1,0)$, $A_* \in [0, \half Z_*]$, $\lambda \geq \frac{3}{2}Z_*$, and $k \leq -1$. Then for each integer $N \geq 0$, there is a unique $\e_N \geq 0$ such that $\mc{W}^-_{\e_N}$ is a saddles connector with winding number $w_{\e_N} = N$. Moreover, they are monotonic: $0 \leq \e_0 \leq \e_1 \leq \dotsb$.
\end{prop}

\proof
Fix $N \geq 0$. By Proposition \ref{prop: large wind num}, there is an $\e''$ such that $w_{\e''} \geq N$.  By Lemma 3.3 in \cite{KLTzGKN}, it follows that $w_{\e} \geq N$ for all $\e > \e''$. 
Therefore we can assume $\e'' > 0$ without loss of generality. Using $\e' = 0$ from Proposition \ref{prop: lower barrier for neg k}, the existence and uniqueness of $\e_N \geq 0$ follows from Proposition \ref{prop: green}. 
The monotonicity follows from another application of Lemma 3.3 in \cite{KLTzGKN}. 
\qed

\medskip

The previous proposition establishes saddles connectors for $k \leq -1$. Next we establish them for $k \geq 1$.

\medskip

\begin{lem}\label{lem: horiz barriers}
Fix $\mathcal{A} \in (0,Z_*]$ and set   $\delta := 1 +\frac{\mathcal{A}^2}{Z_*^2}$.
Let $A_* \in [0, \mathcal{A})$, $\g \in (-\frac{1}{\delta}, 0)$, and $\e = 0$.  
Then for each integer $k \geq 1$, we have
\begin{itemize}
\item[\emph{(a)}] $\dot{\Omega}|_{\Omega = -\frac{\pi}{2}}(\eta) > 0$ for all $\eta \in [0, a_k]$ where $a_k = \frac{\l}{k + \l -\delta\g}$,
\item[\emph{(b)}] $\dot{\Omega}|_{\Omega = -\frac{3\pi}{2}}(\eta) > 0$ for all $\eta \in [b_k, 1]$ where $b_k = \frac{\l}{k + \l +\delta\g}$.
\end{itemize}
\end{lem}

\begin{rem}\hspace{-2pt}
The assumption $\g \in (-\frac{1}{\delta}, 0)$ in the previous lemma is used  only to ensure that $b_k \in (0,1)$ for all~$k$. $\square$
\end{rem}

\proof
We prove (a); the proof of (b) is analogous. 
At $\e =0$, we have
\begin{align*}
\dot{\Omega}|_{\Omega = -\frac{\pi}{2}}(\eta) \,=\,  -\frac{2}{g(\eta)}\big(k\eta(1-\eta) - \lambda (1-\eta)^2 \big) + \frac{2\eta^2}{g^2(\eta)}\g (1-\eta).
\end{align*}
It suffices to show 
\begin{equation}\label{eq: horizontal barrier}
-k\eta +  \lambda (1-\eta) + \frac{\eta^2 \g}{g(\eta)} > 0 \quad \text{ for } \quad \eta \in [0,a_k].
\end{equation}
From eq. \eqref{eq: g(eta)}, we see that the assumption $A_* < \mathcal{A}$ implies that $g^2(\eta) > \frac{\eta^2}{\delta^2}$; this follows since the discriminant of the quadratic $g^2(\eta) - \frac{\eta^2}{\delta^2}$ is negative for $A_* < \mathcal{A}$:
$$
\text{discriminant}\left(g^2(\eta) - \frac{\eta^2}{\delta^2}\right) = 4\bigg(A^2_* + Z^2_*(1-\delta^2)\bigg) < 4\bigg(\mathcal{A}^2 + Z^2_*(1-\delta^2)\bigg) = 0.
$$
Plugging in $g^2(\eta) > \frac{\eta^2}{\delta^2}$ into eq. \eqref{eq: horizontal barrier}  (recall $\g < 0$), we have
$$
-k\eta +  \lambda (1-\eta) + \frac{\eta^2 \g}{g(\eta)} > -k\eta + \lambda(1-\eta) +\delta\g\eta = -\eta(k + \lambda - \delta\gamma) + \lambda \geq 0.
$$
The last step follows since $\eta \leq a_k$.
\qed

\medskip

Consider the line segment in the $(\eta, \Omega)$ plane connecting the points $(a_k, -\frac{\pi}{2})$ and $(b_k, -\frac{3\pi}{2})$ where $a_k$ and $b_k$ are the same as in Lemma \ref{lem: horiz barriers}. 
The next lemma shows that the slope field $\frac{\dot{\Omega}}{\dot{\eta}}$ is larger than the slope of this line (i.e. larger than $-\frac{\pi}{b_k - a_k}$). 
Therefore this line segment joined together with the horizontal line segments in Lemma \ref{lem: horiz barriers} produce a lower barrier.

\medskip

\begin{lem}\label{lem: slanted barrier}
Set $\mathcal{A} =0.1 Z_*$, so $\delta = 1 + \frac{\mathcal{A}^2}{Z_*^2} = 1.01$. 
Set $\nu=3$ and  $\La = \frac{\alphaS}{2\pi}\frac{1}{\nu\delta} \approx 0.000383$. 
Assume $A_* \in [0, \mathcal{A})$, $\g \in (-\frac{1}{\nu\delta}, 0)$, $\l \in [\frac{3}{2}Z_*, \La]$,  and $\e = 0$.  
Then for each integer $k \geq 1$, the slope field satisfies $\frac{\dot{\Omega}}{\dot{\eta}} > -\frac{\pi}{b_k -a_k}$ for all $\eta \in [a_k, b_k]$ and all $\Omega \in [-\pi, \pi)$, where $a_k$ and $b_k$ are the same as in Lemma \emph{\ref{lem: horiz barriers}}.
\end{lem}

\begin{rem} 
The largest value $\La$ for the anomalous magnetic moment term $\l$ is twice what is experimentally observed, so our range for $\l$ agrees with empirical data. $\square$
\end{rem}

\proof
We want to show $\dot{\Omega} + \dot{\eta} \frac{\pi}{b_k - a_k} > 0$. 
For $\e =0$, we have
\[
\dot{\Omega} + \dot{\eta}\frac{\pi}{b_k - a_k} \,=\, A\cos\Omega + B\sin \Omega + C,
\]
where
$$
 A := \frac{2\eta^2}{g(\eta)},\quad
 B := \frac{2}{g(\eta)}\big(k\eta(1-\eta) - \lambda (1-\eta)^2\big),\quad
 C:= \frac{2\eta^2}{g^2(\eta)}\g(1-\eta)+\eta(1-\eta)^2\frac{\pi}{b_k - a_k}.
$$

Compare $A \cos \Omega + B \sin \Omega + C$ with the line $A x + By + C = 0$ in the $(x,y)$ plane. 
The distance from the origin $(0,0)$ to the line is given by $\frac{|C|}{\sqrt{A^2 + B^2}}$. 
Therefore the line never intersects the unit circle provided $C^2 - A^2 - B^2 > 0$. 
(Notice that the dependence on $\Omega$ has vanished; hence these results will be valid for all $\Omega \in [-\pi, \pi)$.) 
If this holds, and we show for some $\Omega_0$ that $A \cos \Omega_0 + B \sin \Omega_0 + C > 0$ for all $\eta \in [a_k, b_k]$, then the unit circle lies in the half of the $(x,y)$ plane where $Ax + By + C > 0$. 
For simplicity, we choose $\Omega_0 = 0$. 
Therefore it suffices to show the following:
\begin{itemize}
\item[(1)] $C^2- A^2- B^2 > 0$ for all $\eta \in [a_k, b_k]$ and all integers $k \geq 1$.
\item[(2)] $A + C > 0$ for all $\eta \in [a_k, b_k]$ and all integers $k \geq 1$.
\end{itemize}

We first show (2). 
Recall from the proof of Lemma \ref{lem: horiz barriers} that $g^2(\eta) > \frac{\eta^2}{\delta^2}$ since $A_* < \mathcal{A}$. 
Therefore
\begin{align*}
\frac{A+C}{1-\eta} \,>\, \frac{C}{1-\eta}\,&>\,\frac{2}{\delta}\g + \eta(1-\eta)\frac{\pi}{b_k - a_k} && \text{since } g^2(\eta) > \frac{\eta^2}{\delta^2},
\\
&>\, \frac{2}{\delta}\g + a_k(1-b_k)\frac{\pi}{b_k - a_k} 
\\
&>\, -\frac{2}{\nu\delta^2} -\frac{\pi}{2} +\frac{\pi \nu}{2}k && \text{since } \gamma > -\frac{1}{\nu\delta},
\\
&>\, 0.
\end{align*}

Now we prove (1). 
From the above, we can find a lower bound for $C^2$. 
For $\eta \in [a_k, b_k]$, we have 
\[
1-\eta \,\geq\, 1-b_k \,=\, 1 - \frac{\l}{k + \l + \delta \g} \,>\, 1 - \frac{\l}{1 + \l - 1/\nu}  \,=\, \frac{\nu-1}{\nu+\l -1} \,\geq\, \frac{\nu - 1}{\nu + \La-1}.
\]
Therefore a lower bound for $C^2$ is
\[
C^2 \,>\, \left(-\frac{2}{\nu\delta^2} -\frac{\pi}{2} +\frac{\pi \nu}{2}k\right)^2\left(\frac{\nu-1}{\nu+\La-1} \right)^2 .
\]
An upper bound for $A^2$ is
\[
A^2 \,=\, \frac{4\eta^4}{g^2(\eta)} \,<\, 4\delta^2 \eta^2 \,\leq\, 4\delta^2b_k^2 \,=\, 4\delta^2\left(\frac{\l}{k+\l-\delta\gamma}\right)^2 \,<\, 4\delta^2\left(\frac{\La}{1+\La}
\right)^2.
\]
An upper bound for $B^2$ is
\begin{align*}
B^2 &=  \frac{4}{g^2(\eta)}(1-\eta)^2\big(k^2\eta^2 -2k\l\eta(1-\eta) + \l^2(1-\eta)^2 \big)
\\
&< \frac{4\delta^2}{\eta^2} \big(k^2\eta^2 -2k\l\eta(1-\eta) + \l^2(1-\eta)^2 \big)
\\
&\leq 4\delta^2\left(k^2  -2k\l \frac{1-b_k}{b_k} +\l^2\frac{(1-a_k)^2}{a_k^2}\right)
= 4\delta^2\big(-4k\delta\g + \delta^2\g^2\big)
\\
&\leq 4\delta^2\left(\frac{4}{\nu}k + \frac{1}{\nu^2} \right).
\end{align*}
Therefore $A^2+B^2 < C^2$ holds if the following inequality holds for all $k \geq 1$.
\[
4\delta^2\left(\frac{\La}{1+\La}
\right)^2 + 4\delta^2\left(\frac{4}{\nu}k + \frac{1}{\nu^2} \right) \,<\, \left(-\frac{2}{\nu\delta^2} -\frac{\pi}{2} +\frac{\pi \nu}{2}k\right)^2\left(\frac{\nu-1}{\nu+\La-1} \right)^2.
\]
Let $u(k)$ and $v(k)$ denote the LHS and RHS, respectively. 
We want to show $u(k) < v(k)$ for all $k \geq 1$.  
$u(k)$ is linear in $k$, while  $v(k)$ is quadratic in $k$. 
Plug in the values $\delta = 1.01$, $\nu = 3$, $\La \approx 0.000383$ into $u(k)$ and $v(k)$. 
At $k =1$, we have $u(1) \approx 5.89$ and $v(1) \approx 6.19$. 
Moreover,  $u'(1) = 16\frac{\delta^2}{\nu} \approx 5.44$, while $v'(1)   \approx 23.4$. Thus $u(k) < v(k)$ for all $k \geq 1$.
\qed

\medskip

It is clear from the proof of the previous lemma that the chosen values for $\mathcal{A}$, $\nu$, and $\La$ are not strict. 
For example, one can increase $\La$ by increasing $\nu$. 

\begin{prop}\label{prop: lower barrier for pos k}
Let $\mathcal{A}$, $\delta$, $\nu$, and $\La$ be as in Lemma \emph{\ref{lem: slanted barrier}}. 
The winding number for $\e = 0$, satisfies $w_0 \leq 1$ for all $A_* \in [0, \mathcal{A}), \g\in (-\frac{1}{\nu\delta},0)$, and $\l \in [\frac{3}{2}Z_*, \La]$.
\end{prop}

\proof
This follows by Lemmas \ref{lem: horiz barriers} and \ref{lem: slanted barrier}. 
The lower barrier is piecewise linear and constructed by the horizontal lines $\Omega = -\frac{\pi}{2}$ for $\eta \in [0,a_k]$ and $\Omega = -\frac{3\pi}{2}$ for $\eta \in [b_k,1]$ along with the line segment connecting the points $(a_k, -\frac{\pi}{2})$ and $(b_k, -\frac{3\pi}{2})$ in the $(\eta, \Omega)$ plane. 
\qed

\medskip

The following proposition is the analogue of Proposition \ref{prop: saddles connector for negative k} for $k \geq 1$.

\medskip

\begin{prop}\label{prop: saddles connector for positive k}
Let $\mathcal{A}$, $\delta$, $\nu$, and $\La$ be as in Lemma \emph{\ref{lem: slanted barrier}}.  
Fix $A_* \in [0, \mathcal{A}), \g \in (-\frac{1}{\nu\delta},0)$, $\l \in [\frac{3}{2}Z_*, \La]$, and $k \geq 1$.  
For each integer $N \geq 1$, there is a unique $\e_N \geq 0$ such that $\mc{W}^-_{\e_N}$ is a saddles connector with winding number $w_{\e_N} = N$. 
Moreover, they are monotonic: $0 \leq \e_1 \leq \e_2 \leq \dotsb$.
\end{prop}

\proof
Use the same argument as in the proof of Proposition \ref{prop: saddles connector for negative k}, but replace Proposition \ref{prop: lower barrier for neg k} with Proposition \ref{prop: lower barrier for pos k}.
\qed

\medskip

For positive integers $k$, the previous proposition establishes winding numbers for all $N \geq 1$. 
From Proposition \ref{prop: upper barrier}, we know that there are no negative-valued winding numbers. 
This leaves open the possibility for the dynamical system to have a winding number of $0$. However, for the usual Coulomb problem on Minkowski spacetime, a winding number of zero is not obtained, see section 7.4 of \cite{ThallerBOOK} (and also section 4 of \cite{KLTzGKN}). 
The next proposition finds sufficient conditions guaranteeing that the dynamical system does not have winding number $0$ when $k$ is positive.

\medskip

\begin{prop}\label{prop: positive wind numbger}
 Let $A_* \in [0, \tfrac{1}{10}Z_*)$, $Z_* \in [\half \alphaS\sqrt{\mc{G}}, \sqrt{\mc{G}})$, and $\lambda \in [\tfrac{3}{2}Z_*,\tfrac{1}{2\pi}\alphaS]$. 
 Then for all $\e \in [0,1)$ and integers $k \geq 1$, the winding number satisfies $w_\e \geq 1$.
\end{prop}

\begin{rem} 
In this proposition, note that $\alphaS \sqrt{\mc{G}} \approx 3.58\cdot 10^{-24}$ is the value of $Z_*$ for when $Z=1$. 
In terms of $\g$, the range $Z_* \in [\frac{1}{2}\alphaS\sqrt{\mc{G}}, \sqrt{\mc{G}})$ is equivalent to $\g \in (-1, \tfrac{1}{2}\alphaS]$. 
In terms of elementary charge units $Z\in\Nset$, the proposition is valid for $1\leq Z \leq 137$. $\square$
\end{rem}

\proof
We want to construct an upper barrier for the system. 
Since $\dot{\Omega}$ decreases as $\e$ increases, it suffices to find an upper barrier for $\e = 0$. Therefore, for the remainder of this proof, we set $\e = 0$. 

By the Hartman--Grobman theorem, the unstable manifold emanating from $S^-$ has tangent vector $\pd_\eta - 2\frac{\l}{Z_*}\pd_\Omega$ at $S^-$. 
Therefore the unstable manifold begins below the horizontal line $\Omega = 0$. 
Now consider the slope field along $\Omega = 0$. We have
\[
\dot{\Omega}|_{\Omega = 0} \,=\, \frac{2\eta^2}{g(\eta)} + \frac{2\eta^2}{g^2(\eta)}\g(1-\eta).
\]
This  is negative provided $g(\eta) < -\g(1-\eta)$, which holds iff $g^2(\eta) < \g^2(1-\eta)^2$, i.e. iff
\[
a\eta^2 + b\eta + c < 0,
\]
where
$$
a := 1 + 2A_* + Z_*^2 - \g^2,\quad
b := -2(A_* + Z_*^2 - \g^2),\quad
c := Z_*^2 - \g^2.
$$
The assumption $Z_* < \sqrt{\mc{G}}$ is equivalent to $\g^2 < 1$. 
Therefore the parabola opens up. 
We see that $a\eta^2 +b\eta + c < 0$ holds for all positive $\eta$ less than 
$$
 \frac{-b + \sqrt{b^2-4ac}}{2a} = \frac{A_*+Z_*^2-\g^2 + \sqrt{A_*^2-Z_*^2+\g^2}}{1+2A_*+Z_*^2-\g^2}
 > \frac{-Z_*^2(\frac{1}{\mc{G}}-1)+Z_*\sqrt{\frac{1}{\mc{G}}-1}}{1+0.2Z_* - Z_*^2(\frac{1}{\mc{G}}-1)}.
 $$
A quick calculation shows that the above expression increases as $Z_*$ increases. 
The minimum value for $Z_*$ we are assuming is one-half the charge of the electron, i.e. $Z_* \geq \half\alphaS\sqrt{\mc{G}}$. 
Plugging this into the above expression gives 
\[
\frac{-b + \sqrt{b^2-4ac}}{2a}  
\,>\,
\frac{-\frac{1}{4}\alphaS^2(1-\mc{G}) +\half \alphaS \sqrt{1-\mc{G}}} {1+0.1\alphaS \sqrt{\mc{G}} - \frac{1}{4}\alphaS^2 (1-\mc{G})}
\,\approx\, 0.00364 \,>\, 0.003.
\]
We conclude that $\dot{\Omega}$ is negative along the horizontal line $\Omega = 0$ for $\eta \in (0, 0.003)$. Set 
\[
\eta_0 \,:=\, \frac{1}{10} \cdot 0.003 = 0.0003.
\]
We form a slanted barrier given by the line segment joining the points $(\eta_0, 0)$ and $(1, -\pi)$ in the $(\eta, \Omega)$ plane. This line segment is given by
\[
\Omega_0(\eta) \,:=\, -\frac{\pi}{1-\eta_0}(\eta - \eta_0).
\]
To show that this line segment is a barrier, we want to show that the slope field $\frac{\dot{\Omega}}{\dot{\eta}}$ is less than the slope of $\Omega_0(\eta)$. 
Therefore we want to show 
\[
\dot{\Omega}\big(\eta, \Omega_0(\eta)\big) + \frac{\pi}{1-\eta_0}\eta(1-\eta)^2 \,<\,0 \quad \text{ for all } \quad \eta \in [\eta_0, 1).
\]
First we estimate $\dot{\Omega}\big(\eta, \Omega_0(\eta)\big)$. By definition, we have
\[
\dot{\Omega}\big(\eta, \Omega_0(\eta)\big) \,=\, \frac{2\eta^2}{g(\eta)} \cos \Omega_0(\eta) + \frac{2}{g(\eta)}\big(k\eta(1-\eta) - \lambda (1-\eta)^2 \big)\sin \Omega_0(\eta) + \frac{2\eta^2}{g^2(\eta)}\g(1-\eta).
\]
We estimate each term separately. In the following, we use the bounds
\[
g(\eta) \,>\, \frac{\eta}{1.01} \quad \text{ and } \quad g(\eta) \,<\, \eta + \sqrt{\mc{G}}.
\]
The first inequality was shown in the proof of Lemma \ref{lem: horiz barriers}. 
The second inequality follows since $g(\eta) < (1-Z_*)\eta + Z_* < \eta + \sqrt{\mc{G}}$.
Using the above bounds, we have
\begin{align*}
\frac{2\eta^2}{g(\eta)}\cos \Omega_0(\eta) \,&=\, \frac{2\eta^2}{g(\eta)}\cos\Omega_0(\eta)\mathds{1}_{[\eta_0,\eta_1]}(\eta) + \frac{2\eta^2}{g(\eta)}\cos\Omega_0(\eta)\mathds{1}_{[\eta_1,1]}(\eta) \\
&<\, 2.02\eta \cos \Omega_0(\eta)\mathds{1}_{[\eta_0,\eta_1]}(\eta) + \frac{2\eta^2}{\eta + \sqrt{\mc{G}}}\cos \Omega_0(\eta)\mathds{1}_{[\eta_1,1]}(\eta).
\end{align*}
Above, $\mathds{1}$ denotes the indicator function and $\eta_1 := \frac{1 + \eta_0}{2} =0.5015$. 
Note that $\cos \Omega_0(\eta)$ is positive on $(\eta_0, \eta_1)$ and negative on $(\eta_1, 1)$ which justifies the above inequality.

Continuing with our estimate of $\dot{\Omega}\big(\eta, \Omega_0(\eta)\big)$, we have
\[
\frac{2}{g(\eta)}k\eta(1-\eta) \sin \Omega_0(\eta) \,<\, \frac{2}{\eta + \sqrt{\mc{G}}}\eta(1-\eta)\sin \Omega_0(\eta),
\]
and
\[
-\frac{2}{g(\eta)}\lambda(1-\eta)^2\sin \Omega_0(\eta) \,<\, -\frac{2.02}{\eta}\frac{1}{2\pi}\alphaS(1-\eta)^2\sin \Omega_0(\eta),
\] 
and
\[
\frac{2\eta^2}{g^2(\eta)}\g(1-\eta) \,<\,-\frac{\eta^2}{(\eta + \sqrt{\mc{G}})^2}\alphaS(1-\eta).\] 
We conclude that for $\eta\in [\eta_0,1]$, we have
\begin{align*}
\dot{\Omega}\big(\eta, \Omega_0(\eta)\big) \,&<\,  2.02\eta \cos \Omega_0(\eta)\mathds{1}_{[\eta_0,\eta_1]}(\eta) + \frac{2\eta^2}{\eta + \sqrt{\mc{G}}}\cos \Omega_0(\eta)\mathds{1}_{[\eta_1,1]}(\eta)
\\
&\:\:\:\:+\frac{2}{\eta + \sqrt{\mc{G}}}\eta(1-\eta)\sin \Omega_0(\eta) -\frac{2.02}{\eta}\frac{1}{2\pi}\alphaS(1-\eta)^2\sin \Omega_0(\eta)
\\
&\:\:\:\:-\frac{\eta^2}{(\eta + \sqrt{\mc{G}})^2}\alphaS (1-\eta)
\\
&=:\, h(\eta).
\end{align*}
Notice that $h(\eta)$ is solely a function of $\eta$ since $\sqrt{\mc{G}}$ and $\alphaS$ are treated as numerical constants in this proof.

Therefore, to show that the line segment $\Omega_0(\eta)$ is a barrier, it suffices to show
\[
j(\eta) \,:=\, h(\eta) + \frac{\pi}{1-\eta_0}\eta(1-\eta)^2 \,<\, 0 \quad \text{ for all } \quad \eta \in [\eta_0,1).
\]
A plot of $j(\eta)$ shows that it is negative for $\eta \in [\eta_0, 1)$. 
To prove this rigorously, we appeal to algebraic techniques. 
First consider the interval $[\eta_0, \eta_1]$. 
Estimating via Taylor series, we have for $\eta \in [\eta_0, \eta_1]$
\begin{align*}
h(\eta) \,\leq\, \tilde{h}(\eta) \,&:=\, 2.02\eta\left(1-\frac{\Omega_0(\eta)^2}{2!} + \frac{\Omega_0(\eta)^4}{4!}\right) 
\\
&\:\:\:\: + \frac{2}{\eta + \sqrt{\mc{G}}}\eta(1-\eta)\left(\Omega_0(\eta) - \frac{\Omega_0(\eta)^3}{3!} \right) -\frac{2.02}{\eta}\frac{1}{2\pi}\alphaS(1-\eta)^2\Omega_0(\eta)
\\
&\:\:\:\: - \frac{\eta^2}{(\eta + \sqrt{\mc{G}})^2}\alphaS (1-\eta).
\end{align*}
Define $\tilde{j}(\eta) := \tilde{h}(\eta) + \frac{\pi}{1-\eta_0}\eta(1-\eta)^2$. 
It suffices to show $\tilde{j}(\eta)$ is negative for $\eta \in [\eta_0,\eta_1]$. 
Clearing the denominator, set 
\[
\tilde{J}(\eta) \,:=\, \eta(\eta + \sqrt{\mc{G}})^2\tilde{j}(\eta).
\]
Then $\tilde{j}(\eta) < 0$ on $[\eta_0, \eta_1]$ iff $\tilde{J}(\eta) < 0$ on $[\eta_0, \eta_1]$. 
Since $\tilde{J}(\eta_1) \approx -0.064 < 0$, it suffices to show $\tilde{J}(\eta)$ has no zeros between $\eta_0$ and $\eta_1$. 
Indeed, $\tilde{J}(\eta)$ is an eighth degree polynomial, and an application of Sturm's theorem \cite[\S 2.2.2]{BPRbook} (which can be implemented algorithmically  via a computer program) shows that $\tilde{J}(\eta)$ has no zeros between $\eta_0$ and $\eta_1$. 
(When applying Sturm's theorem, it is important to work with rational coefficients. 
Therefore the numerical values in $\tilde{J}(\eta)$ have to be bounded by rational numbers.) 
For the interval $[\eta_1, 1]$, one proceeds analogously, but begins the Taylor series at $\Omega = \frac{\pi}{2}$ instead of $\Omega = 0$.
\qed

\medskip

We now combine the results of this section into our main theorem (cf. the main result in \cite{KLTzGKN}). 
The correspondence in this problem, between the winding numbers of saddles connectors and the parameters in the usual spectroscopic notation of the hydrogen spectrum, is the same as the one outlined in section 4 of \cite{KLTzGKN}. 
This correspondence is essential for relating our saddles connectors to the standard hydrogenic orbitals obtainable for the Dirac Hamiltonian on Minkowski space with no anomalous effects included. 
It is also needed for our numerical analysis in the next section. As an example, the saddles connector for $k = -1$ and $N = 0$ corresponds to the $1s_{1/2}$ orbital (which we believe should also be the ground state in our problem, although this still remains to be proven).

\medskip

\begin{thm}\label{thm: main}
Recall $\alphaS$, $\mc{G}$, $A_*$, and $Z_*$ are defined by \eqref{eq:alphaS}, \eqref{eq: A}, \eqref{eq:A_*}, and \eqref{eq:Z_*}, respectively. 
Suppose $Z_* \in (\tfrac{1}{2}\sqrt{\mc{G}} \alphaS, \tfrac{1}{3.03}\sqrt{\mc{G}})$, $A_* \in [0, \tfrac{1}{10}Z_*)$, and $\l \in [\tfrac{3}{2}Z_*, \tfrac{1}{2\pi}\alphaS)$. 
\begin{itemize}
\item Fix an integer $k \leq -1$. For each integer $N \geq 0$, there is a unique $\e_N \geq 0$ such that $\mc{W}^-_{\e_N}$ is a saddles connector with winding number $w_{\e_N} = N$. 
Moreover, they are monotonic: $0 \leq \e_0 \leq \e_1 \leq \dotsb$.
Lastly, there are no saddles connectors with winding number $\leq -1$.

\item Fix an integer $k \geq 1$. 
For each integer $N \geq 1$, there is a unique $\e_N' \geq 0$ such that $\mc{W}^-_{\e_N'}$ is a saddles connector with winding number $w_{\e_N'} = N$. 
Moreover, they are monotonic: $0 \leq \e_1' \leq \e_2' \leq \dotsb$. 
Lastly, for $\e \in[0,1)$, there are no saddles connectors with winding number $\leq 0$. 
\end{itemize}
\end{thm}

\proof
Recall $Z_*$ and $\g$ are related via $\g = -\tfrac{1}{\sqrt{\mc{G}}}Z_*$.
For $k \leq -1$, the result follows by Propositions \ref{prop: saddles connector for negative k} and \ref{prop: upper barrier}. 
For $k \geq 1$, the result follows by Propositions \ref{prop: saddles connector for positive k} and  \ref{prop: positive wind numbger}.
\qed

\medskip

Two remarks are in order. 
First, in terms of atomic numbers, Theorem \ref{thm: main} is valid for $1 \leq Z \leq 45$. 
The nuclear mass number range is valid for $0 \leq A < \tfrac{1}{10}(\tfrac{1}{\sqrt{\mc{G}}}\tfrac{\mEL}{\mPR}\a_S) Z$, which overwhelmingly contains the range $Z \leq A  < 3Z$ for empirically confirmed nuclei. 
The anomalous magnetic moment of the electron, written in units of the Bohr magneton, $\mu_a = -a \frac{e\hbar}{2\mEL c}$, is valid for $3\sqrt{\mc{G}} \leq a < \frac{1}{Z}$; for our values of $Z$ this range contains $a = \tfrac{\a_S}{2\pi}$, the leading contribution derived from QED. 

Second, in Theorem \ref{thm: main}, we only prove the existence of \emph{non-negative} energy eigenvalues. 
This is not because negative energy eigenvalues do not necessarily exist, but rather that the assumption of non-negative energy eigenvalues simplified the proofs of the propositions in this section. 
Negative energy eigenvalues exist as well beyond where our theorem
applies, which is suggested by our numerical analysis in the next section, but a proof of their existence would require a modification to the proofs in this section. 

\vspace{-10pt}
\section{A Numerical Challenge}\label{sec: numerical}
In this section we present the results of some numerical experimentation we have carried out, using the computing platform MATLAB \cite{MATLAB}, regarding the discrete spectrum of the Dirac Hamiltonian for a single electron on a Reissner--Weyl--Nordstr\"om background spacetime that models the electrostatic vacuum outside the point nucleus of a hydrogenic ion, with the electron assumed to have an anomalous magnetic moment. We will point out what we believe is the main challenge in carrying out this task in a mathematically rigorous way, and state a conjecture we believe is worthy of a careful study.

The basic idea for the numerical computation of energy eigenvalues picks up on the approach we took in analyzing the system (\ref{eq: du/dr}), (\ref{eq: dv/dr}), namely, to use the Pr\"ufer transform to reduce the problem to the study of a single nonlinear first-order ODE \eqref{eq: Omega'} on $(0,\infty)$ that depends on the unknown energy eigenvalue $\varepsilon$.  
By Theorem \ref{thm: L^2} this ODE satisfies an asymptotic boundary value problem 
\begin{eqnarray}
\lim_{r\to 0^+}\Omega(r) &= & 0\label{bc0} \\
\lim_{r\to \infty}\Omega(r) & = & -2\pi N - \arccos \varepsilon \label{bcinfty}
\end{eqnarray}
where $N$ denotes the winding number of the corresponding orbit connecting the two saddle-nodes $S^-$ and $S^+_\e$.  
The reason two boundary values can be specified for this first-order equation is that $\varepsilon$ is unknown.  
From the saddle-point nature of these two equilibrium points we in fact know that the solution in question is unstable, so that for any fixed $N\geq 0$ and a generic value of $\varepsilon\in[-1,1]$, the orbit satisfying \eqref{bc0} will either over-shoot or under-shoot its target, depending on whether $\varepsilon$ is smaller or larger than a critical value $\varepsilon_N\in (-1,1)$, which as we have shown, is an energy eigenvalue for the Hamiltonian under study.  
This therefore makes it possible to follow a bisection algorithm in which, starting from an initial guess for $\varepsilon$, the ODE \eqref{eq: Omega'} is solved over a suitably long interval in $r$ and with only condition \eqref{bc0} enforced, and based on whether the final value of this solution is above or below the target \eqref{bcinfty}, the length of the interval containing the eigenvalue is halved repeatedly until it is below a pre-set tolerance.  
\begin{rem}
The utility of the Pr\"ufer transform in finding numerical approximations for energy eigenvalues of spherically-symmetric Dirac Hamiltonians was noted long ago in \cite{ULEHLA1986355}, and used also in \cite{Schmidt}, but this technique has remained relatively under-appreciated in the mathematical physics community.  
The paper \cite{KieTah14a} seems to be the first to realize that the Pr\"ufer method can be extended to the axisymmetric case, which yields a system of {\em two} coupled ODEs with {\em two} unknown parameters. 
This is the approach taken to obtain the numerical results in \cite{KLTzGKN}. 
Other variants of this method have since been employed in related problems \cite{DKTZ2023}. $\square$
\end{rem}
We use the MATLAB ODE solver routine {\tt ode89}, which is an implementation of Verner's ``most robust" Runge--Kutta 9(8) pair with an 8th-order continuous extension \cite{Ver14}. 
The default absolute tolerance for this ODE solver is $10^{-6}$ and the relative tolerance is $10^{-3}$.  
This is an adaptive method: during the integration the step size is adapted so that the estimated error remains below the specified threshold.  
For the bisection algorithm to find the energy eigenvalues, the tolerance we used is around the machine precision, i.e. $10^{-15}$. 

The immediate challenge in numerically solving the initial value problem (\ref{eq: Omega'}--\ref{bc0}) is that the coefficients of the ODE are singular at $r=0$.  
This is true even in the case $\mathcal{G}=0$. 
Thus, for the purpose of numerically solving this ODE, we need to replace \eqref{bc0} with $\Omega(r_0) = 0$ for some very small $r_0>0$. 
Clearly, $r_0$ needs to be larger than the machine precision.  
Here we have used $r_0 = 10^{-6}$.  
On the other hand, the function $f(r)$ appearing in \eqref{eq: Omega'} has the peculiar behavior near $r=0$, as evident from \eqref{eq: asympt f} that, except for an extremely tiny interval of size $r_{\mbox{\tiny res}} := \alphaS \sqrt{\mathcal{G}}\approx 10^{-23}$ around $r=0$, it is practically equal to 1, which is its exact value in the absence of gravity.  
Clearly, $r_{\mbox{\tiny res}}$ is many order of magnitudes smaller than the machine precision $10^{-15}$,  as a result of which, no difference with the case $f \equiv 1$ will be discernible in the computed solutions. 
We could even use an artificial value for the non-dimensional parameter $\mathcal{G}$ that is much larger than its physical value, and still obtain the same result, so long as $r_0>r_{\mbox{\tiny res}}$.  
On the other hand, a much smaller $r_0$ that is comparable to $r_{\mbox{\tiny res}}$ would result in the numerical scheme encountering such large gradients that it would not be able to choose a step size to proceed with the computation.  
It follows that, keeping $r_0$ at its present value, there is a threshold $\mathcal{G}_{\mbox{\tiny res}} \approx 10^{-10}$ such that for any $\mathcal{G} < \mathcal{G}_{\mbox{\tiny res}}$, our numerical way of computing eigenvalues will not be able to distinguish it from $\mathcal{G} = 0$.  
This approach, therefore, does not tell us much about the effect that ``turning gravity on" would have on the spectral lines of the hydrogenic ion in presence of the electron anomalous magnetic moment. 

\begin{figure}[h]
    \hspace{-1truecm}
    \includegraphics[width=16truecm,height=10truecm]{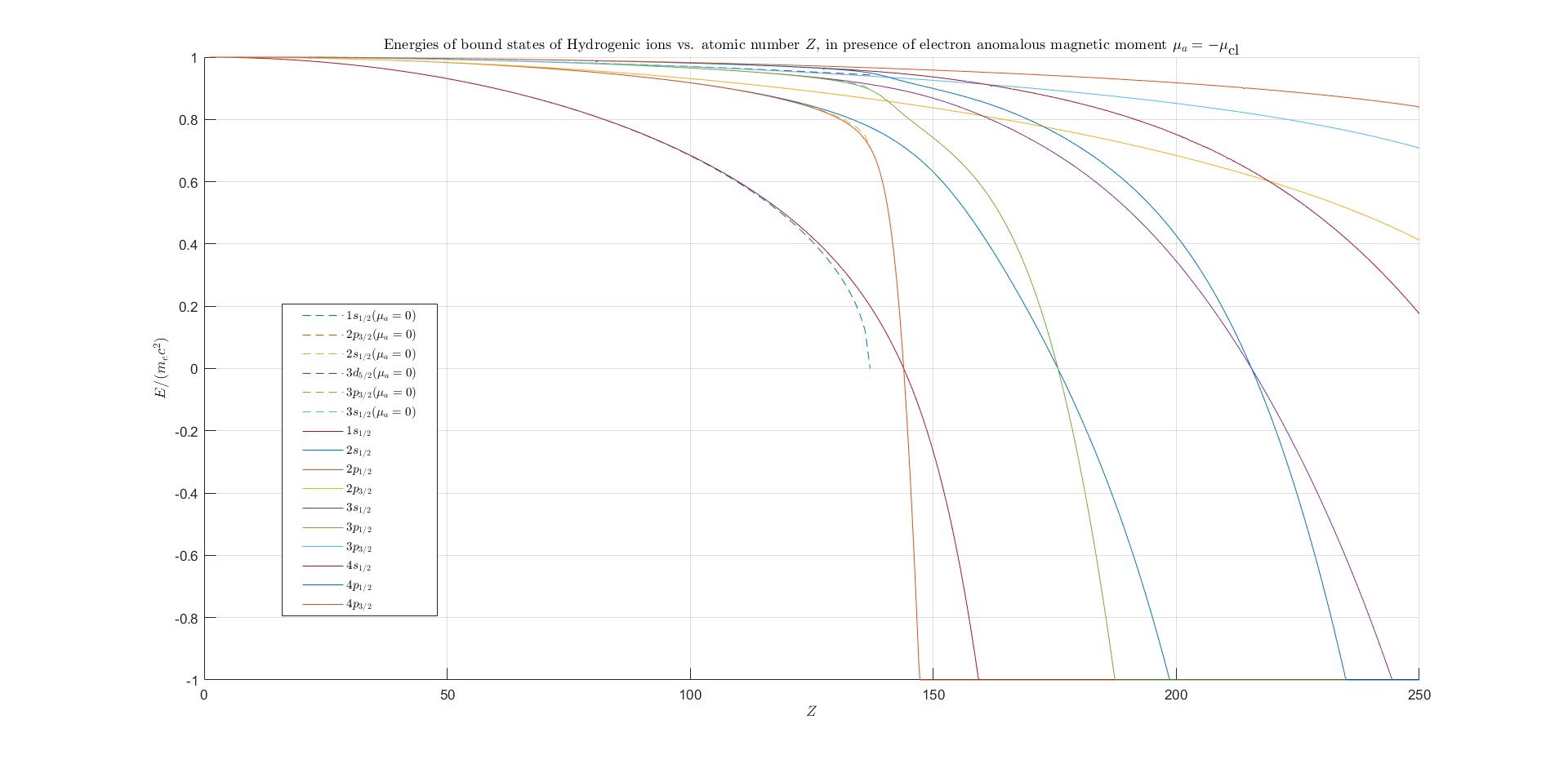}
    \caption{Electron Orbital Energies vs. the Nuclear Charge, with $\mathcal{G} < \mathcal{G}_{\mbox{\tiny res}}$}
    \label{fig:orbitalsnogravity}
\end{figure}

Now, a solution of the asymptotic boundary value problem (\ref{eq: Omega'}), (\ref{bc0}), (\ref{bcinfty}), in addition to the unknown $\varepsilon$ depends on five other constants: the two integers $k$ and $N$, and real numbers $\gamma$, $\lambda$, and $\mathcal{G}$ (through the dependence of $A_*$ and $Z_*$ on $\mathcal{G}$.) 
The dependence of the energy eigenvalues on each of these parameters is of course worthy of study, but from the above discussion it should be clear that the last of these, the dependence on $\mathcal{G}$, presents some unique challenges. 
 The first three parameters, $k$, $N$, and $\gamma$ already feature in the Sommerfeld fine structure formula \eqref{SommerfeldFS}, upon making the substitutions $N := n- |k|$ and $\gamma := -Z\alphaS$.  
 Formally, setting $\lambda = 0$ and $\mathcal{G} = 0$ in the reduced Hamiltonian \eqref{eq: red Hamil K} yields the original Dirac Hamiltonian with pure Coulomb potential, whose energy eigenvalues are given by \eqref{SommerfeldFS}. 
 Plotting the $E_{n,k}$ values in this formula as a function of $Z$ yields a set of curves that are depicted by dashed lines in Figure~\ref{fig:orbitalsnogravity}.  
 The solid lines in this figure are the result of our numerical computation of energy eigenvalues when $\mathcal{G} < \mathcal{G}_{\mbox{\tiny res}}$ and $\lambda$ is set to its physical value. 
Our result here therefore conforms with (and confirms) Fig.~7.1 in \cite{ThallerBOOK}. 

A feature evident in Fig.~\ref{fig:orbitalsnogravity} is that the discrete spectrum continues to exist beyond the range $Z \leq 45$ of $Z$ values for which our main result, Thm.~\ref{thm: main}, holds, and which only proves the existence of positive energy eigenvalues.  
It also appears to be the case, for any $n\geq 1$ and any nonzero $k \in \{-n,-n+1,\dots,n-1\}$, that as $Z$ is increased the energy eigenvalues continue to decrease, eventually crossing the $\varepsilon = 0$ line and becoming negative, and finally touching the continuous spectrum at $\varepsilon=-1$. 
The restrictions in our theorem are therefore far from being optimal. 
Indeed, a rich list of phenomena seems to occur in this system beyond the range of $Z$ corresponding to nuclei so far observed in nature, phenomena that are awaiting to be studied by further careful mathematical analysis of the Hamiltonian \eqref{eq: red Hamil K}.

\vspace{-5pt}
\section{Summary and Outlook}\label{sec: sum and out}

\subsection{Summary}

In this paper we have supplied a complete characterization of the discrete Dirac spectrum of a point electron with fixed anomalous magnetic moment in the Reissner--Weyl--Nordstr{\"o}m spacetime of a point nucleus, though only within certain ranges of the atomic number $Z$, the nuclear mass number $A$, and anomalous magnetic moment $\mu_a$. 
 The $(Z,A)$ range contains approximately 50\% of the empirically confirmed nuclei, from those of hydrogen (H) to those of Rhodium (Rh), and allows us to form a one-to-one correspondence between the eigenfunctions of our Hamiltonian and the orbitals of hydrogen found in textbook quantum mechanics. 
  The discrete spectrum is indexed by two integers $k$ and $N$, where $k$ is the eigenvalue of the spin-orbit operator, and $N$ can be identified with a topological winding number associated with a dynamical system on a compact cylinder that arises from the energy eigenvalue problem of our Hamiltonian. 
 The correspondence to the orbitals of hydrogen is facilitated by defining a principal quantum number $n := N + |k|$.
 
We also carried out a numerical study of the energy eigenvalues as a function of atomic number $Z$; however, we face a challenge: the gravitational coupling constant is so small that our algorithm cannot distinguish between the general and special relativistic problems.

{Our numerical results conform with those in \cite{ThallerBOOK} that, as far as we can tell, nobody had tried to reproduce since their publication.  
 Incidentally, note that the eigenvalues of $H$ \emph{appear} to ``dive into the negative continuum'' when $Z$ becomes sufficiently large. 
  However, by a result of Weidmann \cite{Wei82}, the interior of the essential spectrum of the Dirac Hamiltonian $H$ of an electron with anomalous magnetic moment is \emph{purely absolutely continuous}, and this holds for whether the electron lives in the RWN spacetime of a point nucleus, or in the Minkowski spacetime equipped with a point nucleus. 
Hence, the statement ensuing Fig.~7.1 in \cite{ThallerBOOK}, that the eigenvalues of the special-relativistic $H$ would dive into the lower continuum, should not be taken literally.}

{We emphasize the non-perturbative character of our paper, with its insistence on conceptual clarity that logically leads one to face the spectral questions we have addressed in this paper. 
 In this regard it differs markedly from conventional inquiries into the effects of gravity on the hydrogenic spectra, e.g. \cite{ParkerPimentel}, \cite{Parker}, which only consider the influence of an external gravitational field, and only in a perturbative manner.
In a similar vein, our paper also differs markedly from conventional inquiries into the effect of the electron's (and other leptons') anomalous magnetic moment on the spectra, cf. \cite{GK}, \cite{FanETal}, \cite{CetAL}.}

\subsection{Outlook}

Having a well-defined general-relativistic Dirac operator for hydrogenic ions with nuclei that are treated as  point particles, {and having characterized its discrete spectrum (the essential spectrum had been identified earlier already, in \cite{BMB})}, the litmus test for the operator now is the question whether it has the \emph{correct} energy spectrum, in the sense that its eigenvalue differences reproduce the empirical values of the hydrogen frequencies within an empirically dictated margin of error.
 Since no empirical data exist for $Z>118$, the question is currently decidable only for when $Z\leq 118$. 
    
We emphasize that this question is more difficult to answer than the analogous question for the special-relativistic Dirac operator with the anomalous magnetic moment of the electron taken into account, which implicitly \emph{assumes} that gravitational effects are negligible --- precisely the assumption that {one still has to vindicate rigorously}. 
 Yet, suppose for the sake of the argument right now that gravity is negligible.
 In that case the convergence in the regime $Z\leq 137$ of the {spectrum of the} special-relativistic Dirac operator with $\mu_a$ term to {the spectrum of} the distinguished  self-adjoint extension of the one without it, when $\mu_a\to 0$, as noted in  \cite{KalfSchmidt} and \cite{Schmidt}, is a very helpful piece of information to have. 
 It allows one to work with the explicitly computable spectrum of the distinguished self-adjoint extension of the special-relativistic Dirac operator without $\mu_a$ term, plus $\mu_a$-dependent error bounds, to answer the question. 
{By contrast, when not assuming that gravity is negligible, the analogous strategy does not work for general-relativistic problem with fixed $G>0$}, because one can not take the limit $|\mu_a|\searrow 0$, only the limit $|\mu_a|\searrow \mu_{\rm crit}^{}$, with $\mu_{\rm crit}^{}$ given in (\ref{eq:BMBmu}), but this does not simplify the task at hand. 
 
However, it is reasonable to conjecture that the spectrum of the general-relativistic Dirac Hamiltonian in presence of the anomalous magnetic moment for the electron, converges to the corresponding special-relativistic Dirac spectrum as $\mathcal{G}$ (or equivalently, Newton's constant of universal gravitation $G$) is sent to zero.  
 For reasons described in section 3 and in the Introduction, this conjecture is easy to conceive but (we suspect) not so easy to prove. 
After all, our discussion in section 3 of the RWN metric function $f(r)$ is equally valid if the electron's anomalous magnetic moment is absent from the Dirac Hamiltonian, and in that case one knows through the work of Cohen and Powers \cite{CohenPowers} that the singular behavior of $f(r)$ as $r\to 0$ destroys the essential self-adjointness of the Dirac Hamiltonian with $G=0$ for all $Z\leq 118$; the fact that $f(r)$ is practically 1 except for a tiny vicinity of $r=0$ is thus not a good guide for one's intuition.  
 
Be that as it may, we expect to be able to take the limit $G\searrow 0$ for $\mu_a$ fixed, which --- if the point spectra converge to the special-relativistic $G=0$ spectra --- would reduce the problem to understanding the special-relativistic Dirac spectrum  with anomalous magnetic moment of the electron taken into account, which in turn can be estimated with the help of the strategy mentioned a few lines earlier. 
To rigorously establish the $G\searrow 0$ limit is an interesting problem that we hope to return to in a future work.

Two more open questions concerning the Dirac Hamiltonian of a point electron with fixed anomalous magnetic moment in either the RWN spacetime of a point nucleus, or the Minkowski spacetime equipped with a point nucleus, are suggested by Fig.~1.
  The first question is based on an observation already made in \cite{ThallerBOOK}, namely the apparent level crossings at zero energy. 
Numerically we found that the crossings do not happen exactly at $E=0$, but the difference is so tiny that the deviation from $E=0$ may well be a numerical artifact. 
   It would be good if it could be proved, or disproved, that level crossings happen precisely at $E=0$.
The second question concerns the manner in which the eigenvalue curves meet the negative continuum. 
   Do they meet it with a finite non-zero slope (treating $Z$ as real variable), or with a vanishing slope, or do they never meet the continuum but get asymptotically arbitrarily close to it? 
These are interesting mathematical questions. 
    
{Incidentally, supposing an eigenvalue curve actually meets the negative continuum at $Z=Z_\circ$, say, a natural question to ask in that case is whether $E= - \mEL c^2$ is then part of the point spectrum at $Z_\circ$. (It certainly could not be part of the discrete spectrum, as $E=-\mEL c^2$ belongs to the essential spectrum.)
 That question was already answered in \cite{KieTaZaTopJMP}, though, where a necessary condition was stated for $E= - \mEL c^2$ to be an eigenvalue of the Dirac operator of an electron with anomalous magnetic moment in electrostatic spacetimes of a point nucleus for a vast class of electromagnetic vacuum laws, see Thm.~3.25 in \cite{KieTaZaTopJMP}.
  For the RWN spacetime this condition cannot be met in the naked singularity sector.}

{Another task for the future, motivated by physics,} is to take the finite size of nuclei into account. 
 At large $Z$ values, even in the range of the empirical nuclei, one risks stretching the validity of the point nucleus approximation beyond the estimable.
  Although the charge radius of a $Z=96$ nucleus is only a factor $\approx 7$ bigger than that of a proton (cf. the Nuclear Charge Radii website of the International Atomic Energy Commission), the electron in its ground state is then typically much closer to the nucleus than when $Z=1$, so that the charge distribution of the nucleus should begin to have a significant effect {on the spectra}.

 {A final thought on the problem of essential self-adjointness for hydrogenic ions, related to the finite size of the physical nuclei: It has been argued, cf. \cite{ThallerBOOK}, that the problem of a lack of essential self-adjointness of the Dirac operator for hydrogenic ions (without anomalous magnetic moment of the electron) disappears if the nucleus is equipped with a continuously smeared out charge distribution rather than being treated as a point charge.
 However, this argument misses the point (pardon the pun) that nuclei are spatially extended bound states made of three point quarks, according to the standard model of elementary particle physics.
 Assuming that this standard model captures the nature of nuclei correctly, nuclei should fundamentally be associated with point charges when discussing hydrogenic (and other atomic / ionic) spectra, though the approximation to concentrate all the point charges at a single location is a simplification that turns into an over-simplification when $Z>118$ (as far as essential self-adjointness is concerned; for the spectra it presumably is an over-simplification already for smaller $Z$ values). 
  Yet, since the charges of point quarks have magnitude not more than $\frac23 e$, it is conceivable that essential self-adjointness will hold for Dirac operators of an electron without anomalous magnetic moment in the Coulomb field of nuclei modeled as a sum of Dirac distributions of many point charges with magnitude not more than $\frac23 e$. 
 We are not aware of any work that has addressed the hydrogenic Dirac problem for such models of nuclei.
 It should not be too difficult to address this problem in Minkowski spacetime.
  However, the general-relativistic problem will be formidable.}

{Another issue that needs to be addressed is whether $\mu_a$ changes significantly with the strength of the Coulomb field ($Z$), or the gravitational field ($M$). 
Our use of the classical value for the anomalous magnetic moment independently of $Z$ and of $M$, which is admissible for electrons in weak external fields, may be an oversimplification.
In any event, by using a fixed $\mu_a (= -\mu_{cl})$ for all $Z$ and $M$ values we have followed the tradition \cite{Behncke}, \cite{GesSimTha}, \cite{ThallerBOOK}, \cite{BMB}.} 


\section*{Acknowledgments} 
We thank Moulik Balasubramanian for bringing \cite{ULEHLA1986355} to our attention. We also thank the anonymous referees for a careful reading of our manuscript and for their helpful suggestions.  Eric Ling was supported by Carlsberg Foundation CF21-0680 and Danmarks Grundforskningsfond CPH-GEOTOP-DNRF151.

\medskip
\medskip

\noindent{\bf Declarations.} On behalf of the authors, the corresponding author states that there is no conflict of interest.


\bibliographystyle{amsplain}
\bibliography{references}

\providecommand{\bysame}{\leavevmode\hbox to3em{\hrulefill}\thinspace}
\providecommand{\MR}{\relax\ifhmode\unskip\space\fi MR }
\providecommand{\MRhref}[2]{%
  \href{http://www.ams.org/mathscinet-getitem?mr=#1}{#2}
}
\providecommand{\href}[2]{#2}
\begin{thebibliography}{10}

\bibitem{AKSch}
V.~Arnold, H.~Kalf, and A.~Schneider, \emph{Separated {D}irac operators and
  asymptotically constant linear systems}, Math. Proc. Cambridge Philos. Soc.
  \textbf{121} (1997), 141--146.

\bibitem{Moulik_Thesis}
M.~K. Balasubramanian, \emph{Scalar fields and spin-half fields on mildly
  singular spacetimes}, Ph.D. thesis, Rutgers Univ. (2014), 1--77.

\bibitem{BPRbook}
S.~Basu, R.~Pollack, and M.-F. Roy, \emph{Algorithms in real algebraic
  geometry}, Springer, 2006.

\bibitem{Behncke}
H.~Behncke, \emph{The {D}irac equation with an anomalous magnetic moment},
  Math. Z. \textbf{174} (1980), 213--225.

\bibitem{BMB}
F.~Belgiorno, M.~Martellini, and M.~Baldicchi, \emph{Naked
  {R}eissner--{N}ordstr\"om singularities and the anomalous magnetic moment of
  the electron field}, Phys. Rev. D \textbf{62} (2000), 084014.

\bibitem{CohenPowers}
J.~M. Cohen and R.~T. Powers, \emph{{The general relativistic hydrogen atom}},
  Commun. Math. Phys. \textbf{86} (1982), no.~1, 69 -- 86.

\bibitem{Czarnecki}
A~Czarnecki, \emph{Positronium properties}, Brookhaven Nat. Lab. report
  BNL-HET-99/38 (1999).

\bibitem{CetAL}
A.~Czarnecki, U.~D. Jentschura, K.~Pachucki, and V.~A. Yerokhin,
  \emph{Anomalous magnetic moments of free and bound leptons}, Can. J. Phys.
  \textbf{83} (2005), 1--9.

\bibitem{DarwinDIRACspec}
C.~G. Darwin, \emph{The wave equation of the electron}, Proc. Roy. Soc. London
  \textbf{A118} (1928), 654--680.

\bibitem{DKTZ2023}
S.~Dasgupta, C.~Khurana, and A.~S. Tahvildar-Zadeh, \emph{One-dimensional
  hydrogenic ions with screened nuclear {C}oulomb field}, arXiv:2312.04033
  (2023), 1--25.

\bibitem{QTPDS}
F.~Dumortier, J.~Llibre, and J.~C. Art\'{e}s, \emph{Qualitative theory of
  planar differential systems}, Universitext, Springer-Verlag, Berlin, 2006.

\bibitem{FanETal}
X.~Fan, T.~G. Myers, B.~A.~D. Sukra, and G.~Gabrielse, \emph{Measurement of the
  electron magnetic moment}, Phys. Rev. Lett. \textbf{130} (2023), 071801.

\bibitem{GesSimTha}
F.~Gesztezy, B.~Simon, and B.~Thaller, \emph{On the self-adjointness of {D}irac
  operators with anomalous magnetic moment}, Proc. AMS \textbf{94} (1985),
  115--118.

\bibitem{GordonDIRACspec}
W.~Gordon, \emph{Die {E}nergieniveaus des {W}asserstoffatoms nach der
  {D}iracschen {Q}uantentheorie des {E}lektrons}, Z. Phys. \textbf{48} (1928),
  11--14.

\bibitem{GMRonQED}
W.~Greiner, B.~M\"uller, and J.~Rafelski, \emph{Quantum electrodynamics of
  strong fields}, Springer, 1985.

\bibitem{GK}
S.~Guellati-Khelifa, \emph{Searching for new physics with the electron’s
  magnetic moment}, Physics \textbf{16} (2023), 22.

\bibitem{MATLAB}
The~MathWorks Inc., \emph{Matlab version: 9.13.0 (r2022b)}, 2022.

\bibitem{Jentschura}
U.~D. Jentschura, \emph{Fine-structure constant for gravitational and scalar
  interactions}, Phys. Rev. A \textbf{90} (2014), 022122.

\bibitem{KalfSchmidt}
H.~Kalf and K.~M. Schmidt, \emph{Spectral stability of the {C}oulomb--{D}irac
  hamiltonian with anomalous magnetic moment}, J. Diff. Eq. \textbf{205}
  (2004), 408--423.

\bibitem{KLTzGKN}
M.~K.-H. Kiessling, E.~Ling, and A.~S. Tahvildar-Zadeh, \emph{On the discrete
  {D}irac spectrum of a point electron in the zero-gravity
  {K}err{\textendash}{N}ewman spacetime}, J. Math. Phys. \textbf{63} (2022),
  no.~11, 112301.

\bibitem{KieTah14a}
M.~K.-H. Kiessling and A.~S. Tahvildar-Zadeh, \emph{The {D}irac point electron
  in zero-gravity {K}err-{N}ewman spacetime}, J. Math. Phys. \textbf{56}
  (2015), 042303,1--43.

\bibitem{KieTaZaTopJMP}
M.~K.-H. Kiessling, A.~S. Tahvildar-Zadeh, and E.~Toprak, \emph{On
  general-relativistic hydrogen and hydrogenic ions}, J. Math. Phys.
  \textbf{61} (2020), 092303,1--22.

\bibitem{QED}
D.~Lincoln, \emph{{QED}: Experimental evidence},
  \url{https://www.osti.gov/biblio/1254329}, 4 2016.

\bibitem{Narnhofer}
H.~Narnhofer, \emph{Quantum theory for $\frac{1}{r^2}$ potentials}, Acta Phys.
  Austr. \textbf{40} (1974), 306--322.

\bibitem{NordstromRWN}
G.~Nordstr\"om, \emph{On the energy of the gravitational field in {E}instein's
  theory}, Proc. {K}onigl. {N}ed. {A}kad. {W}et. \textbf{20} (1918),
  1238--1245.

\bibitem{Parker}
L.~Parker, \emph{One-electron atom as a probe of spacetime curvature}, Phys.
  Rev. D \textbf{22} (1980), 1922--1934.

\bibitem{ParkerPimentel}
L.~Parker and L.~O. Pimentel, \emph{Gravitational perturbation of the hydrogen
  spectrum}, Phys. Rev. D \textbf{25} (1982), 3180--3190.

\bibitem{Perko}
L.~Perko, \emph{Differential equations and dynamical systems}, third ed., Texts
  in Applied Mathematics, vol.~7, Springer-Verlag, New York, 2001.

\bibitem{ReissnerRWN}
H.~Reissner, \emph{{\"Uber} die {E}igengravitation des elektrischen {F}eldes
  nach der {E}insteinschen {T}heorie}, Annalen Phys. \textbf{59} (1916),
  106--120.

\bibitem{Schmidt}
K.~M. Schmidt, \emph{Exceptional coupling constants for the {C}oulomb--{D}irac
  operator with anomalous magnetic moment}, J. Comput. Appl. Math. \textbf{194}
  (2006), 17--25.

\bibitem{Schmincke}
U.-W. Schmincke, \emph{Distinguished selfadjoint extensions of {D}irac
  operators}, Math. Z. \textbf{129} (1972), 335--349.

\bibitem{Thaller}
B.~Thaller, \emph{Potential scattering of {D}irac particles}, J. Phys. A: Math.
  Gen. \textbf{14} (1981), 3067--3083.

\bibitem{ThallerBOOK}
\bysame, \emph{The {D}irac equation}, Springer-Verlag, Berlin, 1992.

\bibitem{ThallerREVIEW}
\bysame, \emph{The {D}irac operator. {pp. 23--106 in:} {R}elativistic
  electronic structure theory. {P}art 1. {F}undamentals. {E}d. {P}.
  {S}chwerdtfeger}, Theor. Comp. Chem. \textbf{11} (2002), 1--926.

\bibitem{Vallarta}
M.~Vallarta, \emph{Sommerfeld’s theory of fine structure from the standpoint
  of general relativity}, J. Math. Phys. \textbf{4} (1924), 66--83.

\bibitem{Ver14}
J.~H. Verner, \emph{Explicit {R}unge--{K}utta pairs with lower stage-order},
  Numerical Algorithms \textbf{65} (2014), 555--577.

\bibitem{Weidmann71}
J.~Weidmann, \emph{Oszillationsmethoden f\"ur {S}ysteme gew\"ohnlicher
  {D}ifferentialgleichungen}, Math. {Z}. \textbf{119} (1971), 349--373.

\bibitem{Wei82}
\bysame, \emph{Absolut stetiges {S}pektrum bei {S}turm-{L}iouville-{O}peratoren
  und {D}irac-{S}ystemen}, Math. Z. \textbf{180} (1982), no.~2, 423--427.

\bibitem{Wereide}
T.~Wereide, \emph{The general principle of relativity applied to the
  {Rutherford--Bohr} atom-model}, Phys. Rev. \textbf{21} (1923), 391--396.

\bibitem{WeylRWN}
H.~Weyl, \emph{Zur {G}ravitationstheorie}, Annalen Phys. \textbf{54} (1917),
  117--145.

\bibitem{ULEHLA1986355}
I.~Úlehla and J.~Hořejši, \emph{Generalized {P}rüfer transformation and the
  eigenvalue problem for radial {D}irac equations}, Phys. Lett. A \textbf{113}
  (1986), no.~7, 355--358.

\end{thebibliography}

\end{document}